\documentstyle[12pt]{article}
\newdimen\paperwidth
\newdimen\paperlength
\newdimen\margin
 \newdimen\vmargin
\textwidth=\paperwidth
\advance \textwidth by -2\margin
\advance\hoffset by -1truein
\advance\hoffset by \margin \textheight=\paperlength
\advance \textheight by -2\vmargin
\advance\voffset by -1truein
\advance\voffset by \vmargin
\advance\textheight by
-2\baselineskip
\begin{document}

\begin{titlepage} \title{ {\bf The  Density Matrix Renormalization Group
 }\\
{\bf Method  applied to } \\{\bf Interaction Round a Face  Hamiltonians} }
%\thanks{Work partly supported by CICYT under
%contracts
%PB92-109 , European Community Grant
%ERBCHRXCT920069 } }

\vspace{2cm} \author{ 
{\bf Germ{\'a}n Sierra$^{1}$ and   {\bf Tomotoshi Nishino$^{2}$}  }  \\
$^{1}$ Instituto de Matem{\'a}ticas y F{\'\i}sica Fundamental, 
\\C.S.I.C., Madrid, Spain \\ $^{2}$
Department of Physics, Faculty of Science,
\\ Kobe University, Rokkoudai, Kobe, Japan } 
%\mbox{}
%\vspace{5cm}
%
\date{October 1996} 
\maketitle
\def\baselinestretch{1.3}

\begin{abstract}
Given a Hamiltonian with a continuous symmetry one can generally
factorize that symmetry and consider the dynamics on
invariant Hilbert Spaces. In Statistical Mechanics this procedure
is known as the  vertex-IRF map, and in certain cases, like rotational
invariant Hamiltonians, can be implemented via group theoretical
techniques. Using this map we translate the DMRG method,
which applies to 1d vertex Hamiltonians,
into a formulation adequate to study IRF Hamiltonians. 
The advantage of the IRF formulation of the 
DMRG method ( we name it IRF-DMRG),  is that the dimensions of the
Hilbert Spaces involved in  numerical computations 
are smaller than in the vertex-DMRG, since
the degeneracy due to the symmetry has been eliminated. 
The IRF-DMRG admits a natural and  geometric formulation 
in terms of the paths or string algebras used in Exactly Integrable
Systems and Conformal Field Theory. We illustrate the
IRF-DMRG  method with the study of the SOS model which corresponds 
to the spin 1/2 Heisenberg chain and the RSOS models with Coxeter diagram
of type A, which correspond to the quantum group invariant XXZ chain.

\end{abstract}

\vspace{2cm} PACS numbers: 05.50.+q, 11.10.Gh, 75.10.Jm

%\vskip-17.0cm  \rightline{{\bf September 1996}}
\vskip3in
\end{titlepage}

\newpage
\def\baselinestretch{1.5} \noindent

\section*{Introduction}

The Density Matrix Renormalization Group is a powerful
numerical real space RG method introduced by White in 1992 to study
quantum lattice Hamiltonians of interest in Condensed Matter and
Statistical Physics \cite{W1}. 
This method has its roots in Wilson's 
solution of the Kondo problem \cite{Wi}, but it is not confined to
impurity  problems. The DMRG method overcomes the problems of the old 
Block RG method of the SLAC \cite{SLAC} 
and Paris groups \cite{Paris},
which in many cases gives qualitative correct results but lacks
numerical accuracy ( for a review of the Block RG method see 
\cite{BRG, hT}) . 
The DMRG is well suited for 1d problems as spin chains
\cite{WH, SA} , but it has also
been applied successfully  to  ladder systems \cite{WNS}
and large 2d blocks \cite{W2}. 
Another related developments  are :  
Generalization of the DMRG  to
classical systems and its relation with the Baxter's Corner Transfer
Matrix
\cite{NO},  Variational Formulation of the
DMRG ground state wave function \cite{OR},  Momentum space DMRG 
\cite{X}, application of the DMRG to
transfer matrices \cite{Ni}, DMRG study
of quantum systems  at finite temperature \cite{X2},   
analytic formulation of the DMRG \cite{MS1}. 
The correlation between blocks inherent to the DMRG method has
also been implemented in the old Block RG method in references
\cite{qRG, Role, CBRG}, etc.

The purpose of this paper is to generalize the DMRG to a class of
models commonly
known in Statistical Mechanics as Interaction Round a Face (IRF) 
or more simply Face models \cite{B}. In these models the lattice variables
are labelled by the points (heights) of a graph ${\cal G}$,
and such that heights
located at nearby sites on the lattice must also be nearest 
neighbours on the graph. The SOS models and the RSOS models,
which are a restricted class on the former ones
are the most interesting
examples of IRF models, due to their connection with Integrable
Models \cite{B, ABF}, Affine Lie algebras \cite{Affine},
Towers of Multi-Matrix Algebras \cite{GHJ,GS1}  and  
Conformal Field Theories (CFT) \cite{P1}.
Other important class of models is given by
the vertex models, where the nearby lattice variables are independent, 
although  the Bolztmann weights may satisfy certain 
conservation laws \cite{LW}.
The well known  Heisenberg, t-J  and Hubbard models, 
are Hamiltonian or transfer matrix  versions  
of vertex models, which can be studied using the DMRG. 
It is for the later class of theories that 
applies the standard  DMRG.
We shall propose in this paper to
translate  the "vertex language",  which is used to
formulate the  DMRG,  into "IRF language". This translation process
is suggested by the fact that some models, like the Baxter's 8 vertex,
can be mapped into IRF models 
\cite{B2,Ji2, P2}. Moreover, if the vertex Hamiltonian has
a symmetry described by some group (or quantum group), 
then the vertex-IRF map
consists in the factorization of that symmetry. In symbolical terms we may
write,

\begin{equation}
{\rm IRF} = \frac{{\rm Vertex}}{ {\rm Symmetry}}  
\label{0.1}
\end{equation}

In the case of the Heisenberg model, the factorization of the
rotational symmetry has lead us to
formulate the DMRG in IRF variables. The vertex-IRF map is given
in this example by the tensor product 
decomposition of irreps of $SU(2)$. 
The heights coincide with the irreps of this group.
From (\ref{0.1}) it is clear that an advantage in working with  
IRF variables is that the symmetry present in the vertex Hamiltonian
is factorized, and consequently the dimension of the Hilbert spaces involved
is much lower. For numerical 
purposes this property is also important since it implies a reduction
of the computational complexity of the problem. 
On the other hand the IRF formulation of the DMRG 
is the most natural one to discuss 
its relation with the corner transfer
matrix formalism and CFT.

The organization of the paper is as follows. In section
I we  review the basic concepts and tools of the IRF models.
In section II we introduce the real space renormalization group method 
applied to IRF Hamiltonians. In section III we define the density
matrix for IRF states and use it to propose the 
IRF-DMRG algorithm. In section IV we apply the IRF-DMRG to the 
SOS model, which corresponds to the spin 1/2 Heisenberg model,
and to the  RSOS models 
whose graph is a Coxeter 
diagram of type A, and which are related to the minimal 
models of Belavin, Polyakov and Zamolodchikov \cite{BPZ}. 
In section V we describe the 
vertex-IRF map for Hamiltonians with spin  rotation symmetry, and derive
the IRF-DMRG from the vertex-DMRG. Finally we state our conclusions and 
prospects.

\section*{I)  IRF Models : Basics}

  The IRF models were introduce by Baxter \cite{B}
as Statistical Mechanical models where 
the  variables are defined on the vertices of 
the square lattice while  the interaction 
are defined  on the faces. These kind of models should be distinguished
from the vertex ones, where the variables are located on the edges
and the interaction is defined on the vertices 
where four edges meet \cite{LW}.
In certain cases there will be a deep relationship between these two types of
models, given by a vertex-IRF map.
  The most interesting class of IRF models 
are the so called graph-IRF models ( for a review see \cite{qb}). The 
heights of these models are labelled
by the vertices of a
graph ${\cal G}$. The allowed 
configurations are restricted by  the constraint that 
the lattice variables that are nearest neigbour in the lattice
are also nearest neighbour in the graph ${\cal G}$. 
A characterization of ${\cal G}$ is given by its
incidence matrix $\Lambda_{a,b}$ which is 1 ( reps. 0) if the heights
$a$ and $b$ are connected (resp. disconnected) 
by a link of ${\cal G}$. We assume
that there is at most one link connecting any two points.
The graphs which we shall study in this paper are bipartite,
which means that they can be partitioned into two subgraphs,
say even and odd, so that any point of one subgraph
is connected only to points of the other subgraph.
 A pair of variables $(a,b)$  is said to be admissible if 
$\Lambda_{a,b} = 1$. In this terminology,
all the heights
connected by a link of the square lattice must be admissible.
An important example of IRF models are the RSOS models 
for which  ${\cal G}$
is a ADE Coxeter diagram, which will be studied
in detail in section IV. 
The $A_r$ graph consist of $r$ points 
labelled by $a= 1, 2, \dots, r$.

In  the Hamiltonian or transfer matrix formulation
of IRF models, 
a state of the Hilbert space is described by the ket,

\begin{equation}
|{\bf a}> = |a_0, a_1, \dots , a_{N}>
\label{1.1}
\end{equation}

\noindent
where $(a_i, a_{i+1})$ is an admissible pair. 
There is a geometrical interpretation of the IRF states as paths
on a Brateli diagram, which is constructed by
folding the graph  as in figure 1, and repeating the pattern along the 
"x-axis" \cite{GHJ,O}. A path $\xi$ 
on the Bratelli diagram is a succession of points $\{\xi(i)\}_{i=0}^N $
such that the couple $(\xi (i), \xi (i+1) )$ coincides with a link of 
the diagram (see figure 2). Another important concept is that of 
a plaquette on the Bratelli diagram. A plaquette is the four-tuple,

\begin{equation}
(a,b,c,d) \equiv \left( \begin{array}{ccc}  & d & \\ a & & c \\ & b &  
\end{array} \right), \;\; (a,b) , (b,c), (a,d), (d,c) : {\rm admissible}
\label{1.2}
\end{equation}

\noindent
and can be identified with the elementary squares or plaquettes of the 
Bratelli diagram.

The IRF models can be defined on periodic or open chains.
In this paper we shall concentrate on the later case. 
To define the dynamics of the IRF model
we shall introduce the plaquette operator
$X_i$,  which gives the infinitesimal evolution of an IRF state 
in the neighbourhood of  the $i^{\rm th}$-site of the chain,

\begin{equation}
X_i | \dots, a_{i-1} , a_{i} , a_{i+1}, \dots>
= \sum_{a'_i} R
\left( \begin{array}{lll}  & a'_i & \\ a_{i-1} & & a_{i+1} \\ & a_i &  
\end{array} \right) | \dots, a_{i-1}, a'_i, a_{i+1}, \dots>
\label{1.3}
\end{equation}

\noindent 
$R( a_{i-1}, a_i, a_{i+1}, a'_i)$  
denotes the local Hamiltonian associated to the
plaquette $(a_{i-1}, a_i, a_{i+1}, a'_i)$.  
If $R( a_{i-1}, a_i, a_{i+1}, a'_i)$  
is replaced by a Boltzmann weight then the operators
$X_i$ are those introduced by Baxter in the study of integrable IRF
models. In our case we are working with infinitesimal versions
of these Bolztmann weights and, on the other hand we do not need
to impose any kind of integrability condition, although it could be
interesting to analize a possible interplay between the RG and
integrability.

The Hamiltonian acting on an open chain is defined as,

\begin{equation}
H = \sum_{i=1}^{N-1} X_i
\label{1.4}
\end{equation}

The time evolution produced by (\ref{1.4}) preserves 
the boundary heights $a=a_0$ and $b=a_N$. We shall call ${\cal H}_{a,b}^N$
the Hilbert space expanded by IRF states with these boundary
conditions,

\begin{equation}
{\cal H}_{a,b}^N = \{ |a_0, a_1, \dots, a_N > || a_0 =a , a_N=b \}
\label{1.5}
\end{equation}

Below we shall consider a generalization of this type of 
Hilbert spaces characterized by fixed boundary conditions at the ends.

To finish this section we shall review other applications of IRF 
ideas in the context of Particle Physics, which will help us
to introduce new concepts in the next section.

It is well known the connection between integrable statistical
models and factorizable S-matrix theories \cite{Zamos}. 
For example, the Boltzmann
weights of the 6 vertex model can be conveniently 
identified with the scattering S-matrix of
the solitons of the sine-Gordon theory. 
This kind of interpretation is
also possible for IRF-Bolztmann weights, which may describe
the S-matrix of solitons (or kinks) which connect different vacua.
A soliton say $S_{a,b}(\theta)$ with rapidity $\theta$, is a field
configuration which connects the vacuum $a$ at $x=- \infty$ with the
vacuum $b$ at $ x= + \infty$. Two solitons say $S_{a,b}(\theta_1)$
and $S_{b,c}(\theta_2)$ can meet  at the common vacuum $b$ and 
after a certain time the middle vacuum $b$ can turn  into a new vacuum,
say $d$. The corresponding S matrix for the process 
$S_{a,b}(\theta_1) S_{b,c}(\theta_2) \rightarrow
S_{a,d}(\theta_2) S_{d,c}(\theta_1)$ is described by the 
Boltzmann weight associated to the plaquette
(\ref{1.2}) and rapidity $\theta_1 -\theta_2$.
In this terminology the IRF state (\ref{1.1}) can be interpreted as
a collective state formed by  N solitons 
connecting the vacuum $a_0$ and $a_N$ through
a series of interpolating vacua $a_1, \dots, a_{N-1}$.

There is yet another interpretation of the IRF states.
If we  view the graph ${\cal G}$  
as the target space of a discretized string, then (\ref{1.1}) becomes the
state of an open string $S_{a,b}$ with fixed  boundary 
conditions at the ends. 
As in the case of solitons, the strings may join
and split in various ways according to graph rules.
In the rest of the paper we shall consider as equivalent 
the interpretations of the IRF states as paths
on a Bratelli diagram,
kinks of a field theory and discrete strings (see figure 3),

\begin{equation}
{\rm Path} = {\rm Kink} = {\rm String} 
\label{1.6}
\end{equation}

\section*{II) RG of IRF-Hamiltonians: Generalities}

The basic problem we want to address is the construction of the
ground state and excited states of the IRF Hamiltonian (\ref{1.4}) for
large values of N. The RG method gives an approximate answer 
to this difficult problem. Wilson's strategy
for  the Kondo problem is to 
start out from small chains and grow them by adding site by site, while 
keeping a fixed, and usually large number of 
states, say m,  as "representatives" of the 
whole chain. This method is known as the onion scheme, to be
distinguished from the Wilson-Kadanoff blocking scheme which 
consists in the partition of  
the chain, or 
more generally the lattice, into blocks which
are afterwards  renormalized getting  smaller  lattices.

The Wilsonian growth process applied to an IRF state 
is depicted in figure 4. A string (kink) with states $\ast$ and
$a$ and its ends, "absorbs" a particle (vacuum) in a state
$b$ becoming a new string (kink) with BC's $(\ast,b)$.
Of course
the pair  $(a,b)$ should be admissible for the absortion process to be
possible. The state $\ast$ at the l.h.s. of the string will be kept
fixed in the construction of "longer" strings. The strings will grow
from their  r.h.s.

\subsection*{ String Hilbert Spaces}

Let us call ${\cal H}^S_{\ast, a}$, or more
simply ${\cal H}^S_{a}$, the Hilbert space of the strings
$S$ which have  BC's  $\ast,a$ at their ends. 
An example of ${\cal H}^S_{a}$ is given by the Hilbert space 
(\ref{1.5}), where N measures the length of the string $S$. 
The RG method will lead to the construction of
Hilbert spaces  ${\cal H}^S_{a}$ which are imbedded
into ${\cal H}^N_{a}$ for some N.

The Hilbert space of the
string $S$ with an added  point $\bullet$ on its r.h.s. will be  denoted by
${\cal H}^{S \bullet} _{a}$ and it is given by,

\begin{equation}
{\cal H}^{S \bullet}_{a} = \oplus_{b | \Lambda_{a,b} =1}
{\cal H}^S_{b} 
\label{2.1}
\end{equation}

\noindent
which implies,

\begin{equation}
{\rm dim}{\cal H}^{S \bullet}_{a} = \sum_b \Lambda_{a,b}
\; {\rm dim} {\cal H}^S_{b}
\label{2.2}
\end{equation}

Proceeding as above one can construct longer strings as for example
$S  \bullet \bullet$. All what is needed is the "fusion" 
matrix $\Lambda$. 
For later convenience we give below the basis of the most common
string spaces,

\begin{equation}
\begin{array}{rl}
{\cal H}^{S }_{a} = & \{ |\xi_{a} > \}  \\
{\cal H}^{S \bullet }_{b} = &
\{ |\xi_{a}  \otimes b> | \Lambda_{a,b} =1 \}   \\
{\cal H}^{S \bullet \bullet  }_{c} = & 
\{ |\xi_{a}  \otimes b  \otimes c> |
\Lambda_{a,b}= \Lambda_{b,c} = 1 \} \\
\vdots &    \end{array} 
\label{2.3}
\end{equation}

A generic Hilbert space of the form 
${\cal H}^{S \bullet \stackrel{ n }{ \cdots} \bullet}$ will
be denoted by ${\cal H}^{S, n \bullet}_a$.
The complete Hilbert space of 
a string $S$ plus n points $\bullet$, consists in the direct sum,

\begin{equation}
{\cal H}^{S, n \bullet} =\oplus_a {\cal H}^{S, n \bullet}_a
\label{2.4}
\end{equation}

The total dimension  of ${\cal H}^{S, n \bullet}$
can be computed from eq.(\ref{2.2}),

\begin{equation}
m_{S, n \bullet} \equiv {\rm dim} {\cal H}^{S, n \bullet} = 
\sum_{a_0, \dots, a_n}
\Lambda_{a_0,a_1} \dots \Lambda_{a_{n-1}, a_n} \;  m_{a_n}
\label{2.5}
\end{equation}

\noindent
where
$m_a = {\rm dim}{\cal H}^{S}_{a} $. 
It is important to realize that the sum over heights in 
(\ref{2.5}) may not contain all the heights of the
graph. For example  for a bipartite graph only even or
odd heights will appear at the right 
end of the string ( in this sense the strings associated to
a bipartite lattice can be classified as even or odd).
Thus if the string $S$  
is even (odd) the string plus one point $S \bullet$ 
will be odd (even),
and $S \bullet \bullet $ will again be even (odd), etc.

\subsection*{ String Operators}

We shall  call  string operators   
those operators ${\cal O}^{S, n\bullet}$ which acting on the Hilbert space 
${\cal H}^{S, n \bullet}$, do not change the height located at
right hand end of the combined system $S  \bullet
\stackrel{n}{\cdots}\bullet $.
Their action on the basis (\ref{2.3})
is given as follows,

\begin{equation}
\begin{array}{rl}
{\cal O}^{S } | \xi_a> = & \sum_{\xi'_a} \;
|\xi'_a> < \xi'_a | {\cal O}^S | \xi_a> \\  & \\
{\cal O}^{S \bullet } | \xi_a \otimes b > = &
\sum_c  \sum_{\xi'_c} \;
|\xi'_c \otimes b > 
< \xi'_c \otimes b | {\cal O}^{S  \bullet}
| \xi_a \otimes b> \\ & \\
{\cal O}^{S \bullet \bullet } | \xi_a \otimes b \otimes c > = &
\sum_{d,e}  \sum_{\xi'_e} \;
|\xi'_e \otimes d \otimes c > 
< \xi'_e \otimes d \otimes c | {\cal O}^{S  \bullet \bullet}
| \xi_a \otimes b \otimes c> \\
\vdots &  \end{array}
\label{2.6}
\end{equation}

The matrix elements of the operators 
${\cal O}^{S, n \bullet}$ appearing in (\ref{2.6}) 
will be denoted by,

\begin{equation}
\begin{array}{rl}
< \xi'_a | {\cal O}^S | \xi_a> = &
{\cal O}_{\xi_a}^{\xi'_a}\left( \ast a \right) \\
< \xi'_c \otimes b | {\cal O}^{S  \bullet} | \xi_a \otimes b>= & 
{\cal O}_{\xi_a}^{\xi'_c} \left(
\begin{array}{ccc}   & c &  \\
                    \ast &  & b \\
                     & a &  \end{array} \right) \\
< \xi'_e \otimes d \otimes c | {\cal O}^{S  \bullet \bullet}
| \xi_a \otimes b \otimes c> = & 
{\cal O}_{\xi_a}^{\xi'_e}\left(
\begin{array}{cccc}   &e & d &  \\
                    \ast & & & c \\
                    & a & b & \end{array} \right) \\
\vdots &  \end{array}
\label{2.7}
\end{equation}

\noindent 
and  can be depicted as 
$2(n+1)-$gons, with a special vertex  $\ast$ from which emanate
two thick lines representing  string 
states labelled by $\xi$ (see figure 5).
The remaining $2n$ thin lines connect admissible pairs of
heights. 
The most  important examples of string operators are given by the
Hamiltonians 
${H}^{S,n \bullet }$. However not all the
string Hamiltonians  are independent. Actually, given 
$ {H}^{S \bullet}$ and the "Boltzmann weight" R (\ref{1.3})
one can  build up the remaining 
Hamiltonians ${H}^{S, n \bullet}$ for $n \geq 2$. The first member of the 
later family , namely  ${H}^S $, has to be given independently,
but quite paradoxically it plays little role in the construction.

As as example we give below the matrix representation of ${H}^{S \bullet
\bullet}$,

\begin{eqnarray}
& { H}_{\xi_a}^{\xi'_e}\left(
\begin{array}{cccc}   &e & d &  \\
                    \ast & & & c \\
                    & a & b & \end{array} \right)   &
\label{2.8} \\
& =  { H}_{\xi_a}^{\xi'_e} \left(
\begin{array}{ccc}   & e &  \\
                    \ast &  & b \\
                     & a &  \end{array} \right) 
\delta_{b,d} \; \Lambda_{b,c} 
+ \delta_{a,e} \; \delta_{\xi_a, \xi'_e} \;
R \left( \begin{array}{ccc}   & d &  \\
                    a &  & c \\
                     & b &  \end{array} \right) & \nonumber
\end{eqnarray}

This eq. is depicted in figure 6,
where we show also the construction of 
${H}^{S, 3 \bullet}$.

\subsection*{ The RG-operation}

 The key point of the RG method is the construction 
of the RG-operator $T$  that truncates 
the Hilbert space ${\cal H}^{S \bullet}$ into 
${\cal H}^{S'}$, where $S'$ represents a string with
one more site than the string $S$, i.e.

\begin{equation}
T : \; {\cal H}^{S \bullet} \longrightarrow 
{\cal H}^{S'}
\label{2.9}
\end{equation}

The matrix representation of $T$ and its hermitean conjugate
$T^\dagger$ are given by,

\begin{eqnarray}
& T | \xi_a \otimes b > = \sum_{\xi'_b}  \;
T_{\xi_a}^{\xi'_b} \left( \begin{array}{ccc} \ast & & b \\ & a &
\end{array} \right) | \xi'_b> & \label{2.10} \\
& T^\dagger | \xi'_b  > = \sum_a \sum_{\xi_a} \; 
\bar{T}^{\xi_a}_{\xi'_b} \left( \begin{array}{ccc}  & a &  \\ \ast &  & b
\end{array} \right) | \xi_a \otimes b> & \nonumber
\end{eqnarray}

\noindent 
where

\begin{equation}
\left[ T_{\xi_a}^{\xi'_b} \left( \begin{array}{ccc} \ast & & b \\ & a &
\end{array} \right) \right]^* =
\bar{T}^{\xi_a}_{\xi'_b} \left( \begin{array}{ccc}  & a &  \\ \ast &  & b
\end{array} \right) 
\label{2.11}
\end{equation}

According to (\ref{2.9}) T is a $m_{S'} \times m_{S \bullet}$ matrix.
Except for the first RG-operations we shall always
keep the same number of states describing the 
renormalized system, i.e.  $ m= m_S= m_{S'}$. 
Both $T$ and $T^\dagger$ can be depicted as triangles
with the  special  vertex $\ast$, which is the origin of two thick
edges which symbolize the old and new (renormalized) strings   
( see figure 7). The truncation operator must satisfy
the equation,

\begin{equation}
T T^\dagger = {\bf 1}
\label{2.12}
\end{equation}

\noindent
which guarantees that $T^\dagger T$ is a projection 
operator  which maps 
${\cal H}^{S \bullet}$ into a subspace which is isomorphic to
${\cal H}^{S'}$.

Eq.(\ref{2.12}) reads in components (see figure 8),

\begin{equation}
\sum_a \sum_{\xi_a} 
T_{\xi_a}^{\xi''_b} \left( \begin{array}{ccc} \ast & & b \\ & a &
\end{array} \right) \;
\bar{T}^{\xi_a}_{\xi'_b} \left( \begin{array}{ccc}  & a &  \\ \ast &  & b
\end{array} \right) = \delta_{\xi'_b , \xi''_b} 
\label{2.13}
\end{equation}

\noindent
Given the operators $T $ and $T^\dagger$ we can renormalized
any operator ${\cal O}^{S, n \bullet}$ down  to an operator 
${\cal O}^{S', (n-1) \bullet}$ by means of   the equation,

\begin{equation}
{\cal O}^{S', (n-1) \bullet} = T \; {\cal O}^{S, n \bullet} \; T^\dagger,
\;\;\; (n \geq 1)
\label{2.14}
\end{equation}

\noindent
which in mathematical terms expresses the commutativity of the
following diagram,

\begin{equation}
\begin{array}{rcl} 
{\cal H}^{S',(n-1) \bullet} & 
\stackrel{{\cal O}^{S', (n-1) \bullet}}{\longrightarrow}
& {\cal H}^{S', (n-1) \bullet} \\
T^\dagger \; \downarrow &  &  \uparrow \; T \\
{\cal H}^{S,n \bullet} &
\stackrel{{\cal O}^{S, n \bullet}}{\longrightarrow}
&  {\cal H}^{S, n \bullet} \end{array}
\label{2.15}
\end{equation}

In eqs.(\ref{2.14}) and (\ref{2.15}) the operators 
$T$ and $T^\dagger$ 
act trivially on  the points beyond the closest one
to the string $S$. 
As an example we give below the renormalization  
of $ {\cal O}^{S, \bullet} $ and ${\cal O}^{S, 2 \bullet}$
(see figures 9 and 10 ),

\begin{eqnarray}
& {\cal O}_\xi^{\xi'}(\ast,b)  =
\sum_{a,c,\eta,\eta'}  
T_{\eta'}^{\xi'} \left( \begin{array}{ccc} \ast & & c \\ & b &
\end{array} \right)
{\cal O}_{\eta}^{\eta'}\left(
\begin{array}{ccc}    & c &  \\
                    \ast &  & b \\
                     & a & \end{array} \right) \;
\bar{T}^{\eta}_{\xi} \left( \begin{array}{ccc}  & a &  \\ \ast &  & b
\end{array} \right) & \label{2.16} \\
& {\cal O'}_{\xi}^{\xi'} \left(
\begin{array}{ccc}   & c &  \\
                    \ast &  & b \\
                     & a &  \end{array} \right) =
\sum_{d,e} \sum_{\eta,\eta'} 
T_{\eta'}^{\xi'} \left( \begin{array}{ccc} \ast & & c \\ & e &
\end{array} \right)
{\cal O}_{\eta}^{\eta'}\left(
\begin{array}{cccc}   &e & c &  \\
                    \ast & & & b \\
                    & d & a & \end{array} \right) \;
\bar{T}^{\eta}_{\xi} \left( \begin{array}{ccc}  & d &  \\ \ast &  & a
\end{array} \right) & \label{2.17} 
\end{eqnarray}

In summary we have presented in this section 
a formalism to deal with the renormalization of generic 
IRF Hamiltonians. In the next section we shall explain
the DMRG algorithm to construct the truncation operator
$T$, which will then allow us to carry out explicit computations.

\section*{III) The IRF-DMRG algorithm}

The "standard" RG method to construct the 
operator $T$
applied to IRF models 
consists in the following  two steps: i) diagonalization of the 
Hamiltonian ${\cal H}^{S \bullet}$ and ii) projection to its
lowest energy states. 
This algorithm treats the system $S \bullet$ as isolated from the rest
of points which one adds in posteriori RG steps. 
In other words, the height associated to the point $\bullet$
in $S \bullet$ is fixed to a given value. Imposing fixed boundary
conditions at the ends of the  blocks in the RG method always leads
to bad results. Instead one should consider a combination
of B.C.'s as in \cite{WN}, or impose open B.C.'s 
as in \cite{Role}.

The DMRG is a way to take care of the influence or
correlations of those points
that have not yet been  added to the block. There are various 
DMRG algorithms: infinite system method, finite system method, etc. 
We shall give in this paper 
the IRF version of the infinite system 
algorithm, which is based
on the superblock formed by a string S, 
a point $\bullet$ and another string $S^R$, 
which is the mirror image or reflection of the string $S$ ( see figure 11). 
The dynamics of the "super-string" $S \bullet S^R$ 
involves all allowed heights at the middle point $\bullet$, 
and in that way one is not commited to a particular B.C. on 
$\bullet$. There is an appealing  electrostatic
analogy  to  understand
the role of $S^R$. Let us recall the mirror image 
method which is used to impose Neumann (open) B.C.'s on the 
electrostatic potential. In this sense the mirror string $S^R$
seems to play a similar role, i.e. that of imposing open B.C.'s on $\bullet$.

A basis of the Hilbert space of the 
super-string  $S \bullet S^R$ is given by,

\begin{equation}
{\cal H}^{S \bullet S^R} = \{ 
| \xi_a \otimes b \otimes \eta_c > \;
 || \; \Lambda_{a,b} = \Lambda_{b,c} = 1 \}
\label{3.1}
\end{equation}

The Hamiltonian which  generates the dynamics of the states belonging 
to ${\cal H}^{S \bullet S^R}$ 
can be obtained using the methods of the last section and
it reads (recall figure 6),

\begin{eqnarray}
& { H}_{\xi, \eta}^{\xi', \eta'}\left(
\begin{array}{ccccc}   &a' & b' & c' &  \\
                    \ast & & & & \ast \\
                    & a & b & c& \end{array} \right)   
=  { H}_{\xi}^{\xi'} \left(
\begin{array}{ccc}   & a' &  \\
                    \ast &  & b \\
                     & a &  \end{array} \right) 
\delta_{b,b'} \delta_{c,c'} \; \Lambda_{b,c} \delta_{\eta, \eta'} 
& \label{3.2} \\
& + \delta_{a,a'}   \delta_{\xi, \xi'} \;
R \left( \begin{array}{ccc}   & b' &  \\
                    a &  & c \\
                     & b &  \end{array} \right) 
\delta_{c,c'} \delta_{\eta, \eta'} 
+\delta_{a,a'} \delta_{b,b'} 
\Lambda_{a,b} \delta_{\xi, \xi'} { H}_{\eta}^{\eta'} \left(
\begin{array}{ccc}   & c' &  \\
                    \ast &  & b \\
                     & c &  \end{array} \right) &  \nonumber
\end{eqnarray}

Now we diagonalize this Hamiltonian and select its  ground state
which is called the target state and can be written as,

\begin{equation}
| \psi_0> = \sum_{a,b,c} \sum_{\xi, \eta} 
\psi_{\xi, \eta} (a,b,c) \; | \xi \otimes b \otimes \eta> 
\label{3.3}
\end{equation}

The mirror string $S^R$ plays an auxiliary role in the
construction and we should get rid of it. The DMRG  proposal is 
to construct the reduced density matrix 
 $\rho^{S \bullet}$ of the subsystem $S \bullet$  
by tracing over the states in $S^R$,

\begin{equation}
\rho^{S \bullet} = {\rm Tr}_{ {\cal H}^{S^R}} \;\; | \psi_0>< \psi_0|
\label{3.4}
\end{equation}

\noindent
In the above trace we shall set the height of the middle point the same
for both the ket and the bras, so that the matrix representation
of  $\rho^{S \bullet}$ will be given by,

\begin{equation}
{\rho}_{\xi}^{\xi'} \left(
\begin{array}{ccc}   & a' &  \\
                    \ast &  & b \\
                     & a &  \end{array} \right) 
= \sum_c \sum_{\eta} 
\psi_{\xi, \eta} (a,b,c) \; \psi_{\xi', \eta}^* (a',b,c)
\label{3.5}
\end{equation}

A normalized ground state  
(\ref{3.3}) yields  a properly
normalized density matrix,

\begin{equation}
{\rm Tr}_{ {\cal H}^{S \bullet} }\; \rho^{S \bullet} = 
\sum_{a,b} \sum_{\xi} \; {\rho}_{\xi}^{\xi} \left(
\begin{array}{ccc}   & a &  \\
                    \ast &  & b \\
                     & a &  \end{array} \right) = 1 
\label{3.6}
\end{equation}

The next step is to diagonalize the matrix (\ref{3.5}) 
in the Hilbert space ${\cal H}^{S \bullet}_b$,
for every value of $b$, keeping
the first $m $  eigenstates with highest eigenvalue. 
These states are the most probable ones to
contribute to the ground state of the super-string. 
Finally the matrix T (\ref{2.9}) is  given by these m column vectors.

Eq.(\ref{3.5}) is very similar to Baxter's definition of the corner
transfer matrix (CTM) for IRF models \cite{B}, in the
sense that one traces over the degrees of freedom of half of the system
while keeping the height located at the edge of the "cut" fixed.
This relation between the DMRG and the CTM has  already 
been  pointed out 
in \cite{NO}, and we expect it to hold also for the 
IRF-DMRG.

This ends our presentation of the IRF-DMRG.

\section*{ IV) The IRF-DMRG at work}

We shall apply below the formalism developed in the 
last two sections to study IRF models that can be obtained
by means of vertex-IRF maps of vertex Hamiltonians. This map
will be explained in detail in section V for the case of the SOS
models.

\subsection*{ SOS model (S=1/2)}

The spin chain Hamiltonian of the Heisenberg model
with spin 1/2 reads,

\begin{equation}
H = \frac{1}{2} \sum_i \left( \vec{\sigma}_i \vec{\sigma}_{i+1}
+1 \right) 
\label{4a.1}
\end{equation}

\noindent
where $\vec{ \sigma}_i$ are Pauli matrices acting at the $i^{th}$ site
of the chain. The choice of (\ref{4a.1}) is motivated by the fact that
$ \frac{1}{2}  \left( \vec{\sigma}_i \vec{\sigma}_{i+1}
+1 \right)$ is the permutation operator acting at the sites
$i $ and $i+1$.
The model defined by (\ref{4a.1})
is equivalent to an IRF model whose graph, denoted 
by $A_{\infty}$, consists in the 
semi-infinite chain of fig. 12. The heights $j=0,1/2,1, \dots$,
that label the points of the graph $A_\infty$, 
are in one-to-one correspondence with the 
irreps of the group $SU(2)$. According to fig. 12 the incidence
matrix of $A_\infty$ satisfies

\begin{equation}
\Lambda_{j,j'} = 1 \; \Longleftrightarrow \; |j-j'| = 1/2
\label{4a.2}
\end{equation}

The IRF-Hilbert space associated to a chain with N-sites
is given by the direct sum ( recall (\ref{2.4})),

\begin{equation}
{\cal H}^N = \; \oplus_j \; {\cal H}^N_j
\label{4a.3}
\end{equation}

\noindent
where ${\cal H}^N_{j}$ is the IRF-Hilbert space of all the states
with total spin $j$,

\begin{equation}
{\cal H}^N_j = \{ |j_0, j_1, \dots, j_N >\; || \;  j_0=0, j_N = j,
\;|j_i - j_{i +1}| =1/2 \; \;\; {\rm for} \;  i=0, \dots, N-1 \}
\label{4a.4}
\end{equation}

According to (\ref{4a.4}) the height $\star$ should be identified 
with the identity irrep (i.e. $\star = 0$). Since the graph 
$A_{\infty}$ contains an infinite number of heights this model
is an unrestricted IRF model called solid-on-solid model (SOS). 
This implies in particular that the
Bratelli diagram consists of a pyramid of infinite height as one moves
from the origin of the diagram (i.e.  $\star$ point = vacuum
representation) to the right
hand side ( fig.13).

The dimension of the Hilbert space (\ref{4a.4}) is given by,

\begin{equation}
{\rm dim} \; {\cal H}^N_j = \left( 
\begin{array}{c} N  \\ \frac{N}{2} - j \end{array} \right) -
\left( \begin{array}{c} N  \\ \frac{N}{2} - j -1 \end{array} \right) 
\label{4a.5}
\end{equation}

This formula can be compared with the number of vertex
states of the standard
formulation of the Heisenberg model, 
with a fixed value of the third component of the spin $s^z$
( see section V) ,

\begin{equation}
{\rm dim} \; {\cal V}^N_{ s^z} = 
\left( \begin{array}{c} N  \\ \frac{N}{2} - s^z \end{array} \right)
\label{4a.6}
\end{equation}

For N even (odd) the ground state will belong to the Hilbert space
with $j = 0 \; ( 1/2)$. As can be seen
from eqs.(\ref{4a.5}) and (\ref{4a.6}) it 
is more efficient, for  numerical purposes, to look for the ground
state of the Heisenberg model in the IRF subspaces 
than in the vertex ones ( see table 1).

\begin{center}
\begin{tabular}{|c|c|c|}
\hline
$N$ & dim${\cal H}^N_{j=0}$(IRF) &   dim${\cal H}^N_{s^z=0}$(vertex) \\
\hline
\hline
4 & 2 & 6  \\
\hline
10 & 42 & 252  \\
\hline
20 & 16 796 & 184 756 \\
\hline
24 & 208 012 & 2 704 156 \\
\hline
$N >>1$ & $\sim 1.5956 {2^N}/N^{3/2}$  &
$ \sim 0.7978 {2^N}/N^{1/2}$  \\
\hline
\end{tabular}
\end{center}
\begin{center}
Table 1
\end{center}

From  (\ref{4a.5}) and (\ref{4a.6}) we get the relation,

\begin{equation}
\frac{ {\rm dim} 
{\cal V}^N_{s^z=0}}{  {\rm dim}
{\cal H}^N_{j=0}} = \frac{N}{2} + 1
\label{4a.7}
\end{equation}

\noindent
which is a numerical version of the eq.(\ref{0.1}) and
shows that the difference between the vertex and IRF
formulations persists in the thermodynamic limit.

The constraints (\ref{4a.2})imply
that there are only 6 different "Boltzmann" weights
whose values are given in table 2.

\begin{center}
\begin{tabular}{|c|c|}
\hline
$ \begin{array}{ccc}  & d & \\ a & & c \\ & b &  
\end{array} $ & 
$ R \left( \begin{array}{ccc}  & d & \\ a & & c \\ & b &  
\end{array}\right) $ \\
\hline
\hline
$ \begin{array}{ccc}  & j & \\ j \pm 1/2& & j \mp 1/2 \\ & j &  
\end{array} $ & 1 \\
\hline
$ \begin{array}{ccc}  & j \pm 1/2 & \\ j  & & j  \\ & j \pm 1/2 &  
\end{array} $  & $  \frac{\mp 1 }{2 j + 1}$ \\
\hline
$ \begin{array}{ccc}  & j \pm 1/2 & \\ j & & j\\ & j \mp 1/2 &  
\end{array} $  & $\frac{ \sqrt{2j( 2 j + 2)}  }{2j +1}$  \\
\hline
\end{tabular}
\end{center}
\begin{center}
Table 2: The SOS (S=1/2) Hamiltonian. 
\end{center}

\subsection*{ Numerical Results}

In table 3 we present the data for 
the ground state energy of the 
Heisenberg Hamiltonian (\ref{4a.1}) for  chains of length 
$N= 6 $ up to 24.
In this IRF-DMRG computation we keep  m=12 states, 
which is a rather modest number of states, while
for the vertex-DMRG we keep  a maximum of 12 states
, since it is impossible to 
fix the number of m for vertex-DMRG, due to
the degeneracy based on SU(2) symmetry. This is one of the
advantages of the IRF-DMRG as compared with the vertex-DMRG.
It is clear that at equal  number of retained states,
the IRF method should give  better results than the vertex method.
This expectation is confirmed in table 3.

\begin{center}
\begin{tabular}{|r|l|l|l|}
\hline
N   &       Exact    &        IRF-DMRG &     Vertex DMRG \\
\hline 
6 &  -2.4871542677758 &    -2.4871542677758 * & -2.4871542677758 * \\
8 &  -3.2498651973757 & -3.2498651973757 *&-3.2498651973757 *\\
10&  -4.0160704145657&    -4.0160704145657 *&  -4.0160641768009 \\
12&  -4.7841812656810&    -4.7841812656810 *&  -4.7835746471807\\
14&  -5.5534493237243&    -5.5534493236562&  -5.5527041895949\\
16&  -6.3234742911502&    -6.3234742887647&  -6.3246392674964\\
18&  -7.0940221370730&    -7.0940221268443&  -7.0953356164320\\
20&  -7.8649466687979&    -7.8649466378157&  -7.8663701424885\\
22&  -8.6361517519671&    -8.6361516790424&  -8.6375643759289\\
24&  -9.4075715208191&    -9.4075713713902&  -9.4089821742785\\
\hline
\end{tabular}
\end{center}
\begin{center}
Table 3: Ground state energy of the Hamiltonian
(\ref{4a.1}). The data followed by "*" are exact.
\end{center}

If we increase the number $ m$ of states retained,
the results converge exponentially fast both in the
vertex-DMRG and IRF-DMRG methods ( fig.14).

\subsection*{ RSOS models}

An interesting generalization of the spin 1/2 Heisenberg chain is
provided by the XXZ Hamiltonian with boundary terms \cite{ABBBQ} ,

\begin{equation}
H^{XXZ} = \frac{1}{2} \left[  \sum_{i=1}^{N-1} \left(
{ \sigma}^X_i {\sigma}^X_{i+1} 
+{ \sigma}^Y_i {\sigma}^Y_{i+1} 
+ \frac{ q + q^{-1}}{2} 
{\sigma}^Z_i {\sigma}^Z_{i+1} \right)
+ \frac{q - q^{-1}}{2} ( { \sigma}^Z_1 - {\sigma}^Z_{N} ) \right]
\label{4b.1}
\end{equation}

\noindent
where $q  = e^{ {\rm i} \gamma} $ is a phase. This Hamiltonian
has very interesting properties: 

\begin{itemize}

\item  The eigenenergies of the N=2M site XXZ chain 
(\ref{4b.1})   coincide with 
those a M-site  self-dual Q-state 
Potts model  with $Q = q + q^{-1}= 2 \; {\rm cos \gamma}$ 
\cite{H,ABB}.

\item  Invariance under the action of the quantum group 
$SU(2)_q$ \cite{PS}.

\item Using q-group theory one can map the vertex Hamiltonian
(\ref{4b.1}) into a RSOS Hamiltonian whose graph is given by the
Coxeter diagram $A_r$ ( see figure 15) \cite{P2}.

\item $H^{XXZ}$  is critical \cite{ABBBQ}
and for $\gamma= \frac{\pi}{r+1}$ it 
belongs  to the universality class of the
minimal CFT's \cite{BPZ} 
with a value of the central charge given by,

\begin{equation}
c = 1 - \frac{6}{r ( r+1)}
\label{4b.2}
\end{equation}

\end{itemize}
We shall study below the RSOS version of the vertex Hamiltonian 
(\ref{4b.1}). A way to arrive to this version consists in writing
(\ref{4b.1}) as follows \cite{TL} ,

\begin{eqnarray}
& H^{XXZ}= \sum_{i=1}^{N-1} \left( \frac{ q + q^{-1}}{4} - e_i \right) & 
\label{4b.3} 
\end{eqnarray}

\noindent
where $e_i$ are the Temperley-Lieb-Jones (TLJ) operators which act
at the positions $i^{th}$ and $(i+1)^{th}$ of the chain and satisfy
the TLJ algebra \cite{qb},

\begin{eqnarray}
& e^2_i = ( q + q^{-1}) e_i & \nonumber \\
& e_i \; e_{i \pm 1} \; e_i = e_i & \label{4b.4} \\
& e_i \; e_j = e_j \; e_i ,\;\;\;\; |i-j| \geq 2 & \nonumber
\end{eqnarray}

\noindent
In the vertex basis   the TLJ operator $e_i$
can be written as follows,

\begin{equation}
e_{i} = {\bf 1}_1 \otimes 
\cdots \otimes {\bf 1}_{i-1} \otimes \left( \begin{array}{cccc} 
                                   0 & 0 & 0 & 0 \\
                                   0 & q^{-1} & -1 & 0 \\
                                   0 & -1 & q & 0 \\
                                   0 & 0 & 0 & 0 \end{array} \right)
\otimes {\bf 1}_{i+2} \cdots \otimes {\bf 1}_N
\label{4b.5}
\end{equation}

The existence  of a vertex-IRF map, and the fact that 
the operators $e_i$  commute with the action of $SU(2)_q$
imply that they can be given 
a representation on the RSOS-Hilbert spaces 
of the face model defined by
the graph $A_r$,

\begin{equation}
e_i | \dots, a_{i-1} , a_{i} , a_{i+1}, \dots>
= \sum_{a'_i} e
\left( \begin{array}{lll}  & a'_i & \\ a_{i-1} & & a_{i+1} \\ & a_i &  
\end{array} \right) | \dots, a_{i-1}, a'_i, a_{i+1}, \dots>
\label{4b.6}
\end{equation}

\begin{equation}
e\left( \begin{array}{ccc}  & d & \\ a & & c \\ & b &  
\end{array} \right) = \delta_{a,c} \; \frac{ \sqrt{ t_b t_d}}{ t_a}
\label{4b.7}
\end{equation}

\noindent
where $t_a$ are the components of
the Perron-Frobenius vector of the incidence matrix
of the graph $A_r$, which are given by

\begin{equation}
t_a = {\rm sin}\left( \frac{ \pi a}{ r+1} \right),  \;\;
a= 1, 2, \dots, r 
\label{4b.8}
\end{equation}

Notice that $t_a$ satisfies,

\begin{equation}
t_{a+1} + t_{a-1} = 2 {\rm cos}\left( \frac{{\pi}}{r+1}\right) \; t_a
\label{4b.9}
\end{equation}

Recalling that the incidence matrix satisfies in this case

\begin{equation}
\Lambda_{a,b} = 1 \Longleftrightarrow |a-b| = 1
\label{4b.10}
\end{equation}

\noindent
one gets that there are only
6 types of "Boltzmann weights", whose expression, 
given in table 4,  can be 
computed using eqs (\ref{4b.3}) and (\ref{4b.7}),

\begin{center}
\begin{tabular}{|c|c|}
\hline
$ \begin{array}{ccc}  & d & \\ a &   & c \\ & b &  
\end{array} $ & 
$ R \left( \begin{array}{ccc}  & d & \\ a &  & c \\ & b &  
\end{array}\right) $ \\
\hline
\hline
$ \begin{array}{ccc}  & a & \\ a \pm 1&  & a \mp 1 \\ & a &  
\end{array}$  & $\frac{{\rm cos}{\gamma}}{2}$ \\
\hline
$ \begin{array}{ccc}  & a \pm 1 & \\ a & & a \\ & a \pm 1 &  
\end{array} $  &$ \frac{{\rm cos}{\gamma}}{2}
- \frac{t_{a \pm 1} }{t_a}$ \\
\hline
$ \begin{array}{ccc}  & a \pm 1 & \\ a &  & a \\ & a \mp 1 &  
\end{array} $  & 
$- \frac{ \sqrt{ t_{a+1} t_{a-1} } }{ t_a }$  \\
\hline
\end{tabular}
\end{center}
\begin{center}
Table 4: The RSOS Hamiltonian for the face model $A_r$. 
\end{center}

In the limit where $r \rightarrow \infty$ the RSOS model $A_r$
becomes  equivalent to the SOS model with graph $A_{\infty}$
studied previously. One can check that the R-matrix
given in table 4 is, up to a constant and a change of basis, the
same as the R-matrix
given in table 2, with the identification $a= 2j+1$.

\subsection*{Numerical Results}

In table 5 we give the ground state energy  per 2 sites 
$E_0(M)/M ( M = N/2)$ of the XXZ  model (\ref{4b.1}), which
coincides with the ground state energy per site 
of the corresponding
Potts model.  This table should be compared
with table 1b in reference \cite{ABBBQ}, 
which was obtained using the Bethe
ansatz. The authors of \cite{ABBBQ} give their results up to
6 decimals ( ours is 9) and the agreement in the
energies holds until the $6^{\rm th}$
digit. The number of states  retained in our computation
is $m=160$.

Using the IRF-DMRG data we have  computed the finite size corrections
to the ground state energy, which are governed by the 
formula \cite{Cardy,Aff},

\begin{equation}
E_0(M)/M = e_{\infty} + \frac{f_\infty}{M} - 
\frac{\pi \varsigma c}{24 M^2} + o(M^{-2})
\label{4b.11}
\end{equation}

\noindent
where $e_\infty$ and $f_\infty$ are, respectively, the bulk
and surface energy per site. $\varsigma$ can be identified
with the spin wave velocity and it is given for the
Potts model by,

\begin{equation}
\varsigma = \frac{ \pi {\rm sin} \gamma}{ 2 \gamma}
\label{4b.12}
\end{equation}

We have used Sach's formula to get the values of the central
charge c \cite{van}.

The outcome of this
computation  is that the IRF-DMRG method
reproduces rather accurately the results obtained using the 
Bethe ansatz. This supports the hypothesis that the DMRG is
in fact an exact numerical RG method.

\begin{center}
\begin{tabular}{|r|c|c|c|c|}
\hline
N/2 &    r=3 (Q=2)    &  r=5 (Q=3) &      r=7 (Q=3.414) & r=inf (Q=4)\\
\hline
2& -1.320 899 500& -1.478 675 250&-1.537 532 848&-1.616 025 403 \\
4& -1.459 958 153& -1.580 754 400&-1.626 304 313&-1.687 466 299\\
8&  -1.532 472 880&-1.636 075 843&-1.675 237 896&-1.727 934 286\\
16&  -1.569 617 420&-1.665 066 291&-1.701 130 752&-1.749 664 452\\ 
32&  -1.588 433 915&-1.679 941 590&-1.714 488 941&-1.760 960 537\\
64&  -1.597 906 426&-1.687 482 175&-1.721 280 643&-1.766 733 433\\
128&  -1.602 659 177&-1.691 279 439&-1.724 706 257&-1.769 649 934\\ 
256&  -1.605 039 733&-1.693 185 000&-1.726 426 767&-1.771 116 470\\
512&  -1.606 231 063&-1.694 139 540&-1.727 288 989&-1.771 851 868\\
$e_\infty$&  -1.607 423 097 & -1.695 095 264& -1.728 152 544&-1.772 588 719 \\
(AB$^3$Q)& -1.607 423 &   -1.695 095&     -1.728 152&    -1.772 588 \\
$f_\infty$ &  0.610 501 838 &0.489 636 433 & 0.442 486 891& 0.377 747 124 \\
(AB$^3$Q)&0.610 502     &0.489 637    &  0.442 487&     0.377 649\\
c &    0.499 942 &   0.798 817&  0.893 150  &   1.038 18\\
(AB$^3$Q)&  0.500 00(1)& 0.799 9(2) &   0.89(3)        &   0.99(2)   \\
(exact)& 0.5 & 0.8 & 0.892 857 & 1 \\
\hline
\end{tabular}
\end{center}
\begin{center}
Table 5: Ground state energy per 2-sites of the RSOS chain.
We also give the results of ref \cite{ABF} (AB$^3$Q) for 
$e_{\infty}$, $f_{\infty}$ and $c$.
\end{center}

\section*{V) Vertex-IRF Map}

We shall illustrate this map in the case of a spin-s chain whose
dynamics is dictated by a 
rotational invariant Hamiltonian. 
The Hilbert space of the spin-s chain with N sites 
will be  denoted by ${\cal V}^{N,s}$ and 
consists in the tensor product of N copies of the 
vector space $V_s = {\bf C}^{2s+1}$, 
where acts the local spin operators
${\bf S}_i$ ( i = 1, \dots , $N$). 
The vertex-IRF map is based on the tensor product decomposition
of the space ${\cal V}^{N,s}$ into its irreducible components.

\subsection*{Vertex Hilbert Spaces $\rightarrow$ IRF Hilbert Spaces}

Using the Clebsch-Gordan decomposition of tensor product
of irreps of $SU(2)$ one can write,

\begin{equation}
{\cal V}^{N,s} = \sum_{0 \leq j \leq 2 s N} {\cal H}^{N,s}_j  \otimes V_j
\label{5.1}
\end{equation}

\noindent
where
${\cal H}^{N,s}_j$ is the generalization of the 
IRF Hilbert space (\ref{4a.4}) to the spin s case. The 
heights  $a_i \in \frac{1}{2} {\bf Z_+} $ that label the 
IRF states are subject to the
following constraints,

\begin{equation}
\begin{array}{ccc}  &  a_0 = 0 & \\ 
& a_1 = S &  \\
|S- a_i|& \leq a_{i+1} \leq & |S + a_{i}| ,\;\; i = 1, \dots, N-1 \\ 
& a_N = j &  \end{array} 
\label{5.2}
\end{equation}

The dimension of ${\cal H}^{N,s}_j$ is given by the number of times
the spin-j irrep appears in the CG-decomposition of
the tensor product $s \otimes
\stackrel{N}{\cdots} \otimes s$,

\begin{equation}
{\rm dim} {\cal H}^{N,s}_j = {\rm multiplicity}\; {\rm of } 
\; V_j \;\;{\rm in} \; {\cal V}^{N,s}
\label{5.3}
\end{equation}

\noindent
${\rm dim} {\cal H}^{N,s}_j $ can be computed using the following
formula,

\begin{equation}
{\rm dim} {\cal H}^{N,s}_j = {\rm dim} {\cal V}^{N,s}_{j}- 
{\rm dim} {\cal V}^{N,s}_{j+1}
\label{5.4}
\end{equation}

\noindent
where $ {\cal V}^{N,s}_{s^z}$ denotes the subspace of $ {\cal V}^{N,s} $ with
a fixed value $s^z$ of the third component of the spin. Eq.(\ref{5.4})
says  that the highest weights with total spin $j$ are given by
the states  with spin $s^z =j$ minus the ones that 
can be obtained from $ {\cal V}^{N,s}_{s^z+1}$ by the
lowering operator $S^-$. The counting of states with a fixed value
of $s^z$ is easily done using the "Bethe method" of starting with
the ferromagnetic state with all the spins up and lowering the spin.
Below we give the formulae  for s=1/2 and 1.

\begin{eqnarray}
& {\rm dim} {\cal V}^{N,s=1/2}_{s^z} = 
\left( \begin{array}{c} N \\ \frac{N}{2} - s^z \end{array} \right)
 & \label{5.5} \\
& {\rm dim} {\cal V}^{N,s=1}_{s^z}=
\sum_{k=0}^{ [(N - s^z)/2 ]}    
\left( \begin{array}{c} N \\  s^z + k  \end{array} \right) \;
\left( \begin{array}{c} N- s^z- k \\ k \end{array} \right)
& \nonumber
\end{eqnarray}

\noindent
where the symbol $[x]$ appearing in the upper
limit of the sum denotes the integer part of x.
The relation between the vertex basis
of the spaces ${\cal V}^{N,s}, V_j  $  and the IRF basis of
${\cal H}^{N,s}_j$ can be obtained using the Clebsch-Gordan
coefficients as follows,

\begin{eqnarray}
&\xi({\bf a}) \otimes e^j_m & \nonumber \\ 
&  = \sum_{m_1, \dots, m_N} 
\left[ \begin{array}{lll}  0 & s & a_1 \\
                           0 & m_1 & n_1 \\
       \end{array} \right]
\left[ \begin{array}{lll}  a_1 & s & a_2 \\
                           n_1 & m_2 & n_2 \\
       \end{array} \right]
\left[ \begin{array}{lll}  a_2 & s & a_3 \\
                           n_2 & m_3 & n_3 \\
       \end{array} \right] \cdots  & 
\label{5.6} \\
& \cdots \; 
\left[ \begin{array}{lll}  a_{N-2}  & s & a_{N-1} \\
                           a_{N-2} & m_{N-1} & n_{N-1} \\
       \end{array} \right]
\left[ \begin{array}{lll}  a_{N-1} & s & a_N \\
                           n_{N-1} & m_N  & n_N \\
       \end{array} \right]
\;\; {\rm e}^s_{m_1} \otimes \cdots \otimes  {\rm e}^s_{m_N}
 & \nonumber 
\end{eqnarray}

\noindent
where $n_i = m_1 + \cdots + m_i $ , $m= n_N = 
\sum_{i=1}^N m_i $, and ${\bf a}$ denote the IRF labels
which satisfy conditions (\ref{5.2}).
A graphical representation of eq.(\ref{5.6}) is given
in Fig.16. The 0 at the upper left of the diagram 
can be identified with the $\star$ symbol introduced in section II.
The vertex-IRF map, as defined by eq.(\ref{5.6}), is nothing but a 
change of basis from vertex variables to IRF ones which achieves
the  factorization of the $SU(2)$ symmetry.

\subsection*{ Vertex-Hamiltonians $\rightarrow$ IRF-Hamiltonians}

The most general form of a rotational invariant Hamiltonian $H$ acting in
${\cal V}^{N,s}$, i.e.

\begin{equation}
[ H , {\bf S} ] = 0 , \;\; {\bf S} = \sum_{i=1}^N {\bf S}_i
\label{5.7}
\end{equation}

\noindent
which is translational invariant and contains 
only nearest neaghbours couplings is given by,

\begin{equation}
H = \sum_{i=1}^{N-1}  \sum_{r=0}^{2s}  \alpha_r \left( {\bf S}_i \cdot
{\bf S}_{i+1} \right)^r  
\label{5.8}
\end{equation}

\noindent
where $\alpha_r$ is a set of 2s+1 coupling constants  
($\alpha_0$ can be put equal to zero since it multiplies the identity
operator). Since (\ref{5.8}) commutes with $SU(2)$ 
it means  that its action affects  only the IRF spaces ${\cal H}^N_j$.
Using group theoretical methods we get,

\begin{eqnarray}
& H \; \xi( {\bf a}) = \sum_{i=1}^{N-1} \;
R
\left( \begin{array}{ll} a_{i-1} & a'_i \\ a_i & a_{i+1} 
\end{array} \right) \xi(\cdots, a_{i-1}, a'_i, a_{i+1}, \cdots) &
\label{5.9}
\end{eqnarray}

\noindent
where the IRF "weights" $R$ can be computed in terms of the
coupling constants $\alpha_r$ and the 6j-symbols 
as follows (the details of this computation will be given elsewhere),

\begin{equation}
R\left( \begin{array}{ll} a_{i-1} & a'_i \\ a_i & a_{i+1} 
\end{array} \right) = \sum_{0 \leq j \leq 2s} 
A^{ss}_j
\left\{  \begin{array}{lll} s & s & j \\
                            a_{i-1} & a_{i+1} & a_i \end{array} \right\}
\;\left\{  \begin{array}{lll} s & s & j \\
                               a_{i-1} & a_{i+1} & a'_{i} \end{array} \right\}
\label{5.10}
\end{equation}

\begin{eqnarray}
&A^{ss}_j = \sum_{r=0}^{2s} \alpha_r  x_j^r & \label{5.11} \\
& x_j = \frac{1}{2} j(j+1) - s (s+1) & \nonumber
\end{eqnarray}

As an example we may choose H to be the sum of all the permutation
operators between nearest neighbours, in which case 
$A_j^{ss}$ turns out to be  a sign factor,

\begin{equation}
H = \sum_i P_{i,i+1} \Longrightarrow  A^{ss}_j = (-1)^{2s-j}
\label{5.12}
\end{equation}

The IRF hamiltonian corresponding to (\ref{5.12}) is given by,

\begin{equation}
R\left( \begin{array}{ll} a_{i-1} & a'_i \\ a_i & a_{i+1} 
\end{array} \right) = 
(-1)^{a_{i-1}+ a_{i+1} - a_i - a_i'}
\left\{  \begin{array}{lll} s & a_{i-1} & a_i \\
                            s & a_{i+1} & a'_i \end{array} \right\} 
\label{5.13}
\end{equation}

The s=1/2 Heisenberg Hamiltonian (\ref{4a.1}) is precisely 
of the form (\ref{5.12}) so that table 2 can be derived
from (\ref{5.13}). In a subsequent publication we shall use eq.(\ref{5.10})
to study higher spin Heisenberg chains in the IRF formalism.

\subsection*{Vertex-DMRG $\rightarrow$ IRF-DMRG}

The DMRG algorithm 
to renormalize a block $B$ plus one point $\odot$  
into a new block  $B'$, 
is based on the superblock
$B \odot \odot B^R$, where $B^R$ is the reflection of the
block B ( we use $\odot$ to distinguish  vertex-points 
from  IRF-points which were denoted above by $\bullet$ ). 
The main steps of the vertex-DMRG are: 

\begin{itemize}

\item Diagonalization of the superblock Hamiltonian
      to find the ground state $|\psi>$. 

\item Construction of the reduced density matrix by tracing over the
states in $\odot B^R$,

\begin{equation}
\rho^{B \odot} = {\rm Tr}_{\odot B^R}  \;\; |\psi_0> < \psi_0| 
\label{5.14}
\end{equation}

\item Diagonalization of $\rho^{B \odot}$ to find the eigenvalues
$w_\alpha$ and eigenvectors $|u^\alpha>$. Discard all but the largest
$m$ eigenvalues and associated eigenvectors.

\item Construct the operator  $T$ using the truncated
eigenvectors $|u^\alpha>$.

\item Renormalize all the operators using the analog of eq. (\ref{2.14})
in the vertex case.

\item Repeat the process for the new block $B'$.

\end{itemize}

These set of rules define the  vertex-DMRG algorithm, 
which applies directly to systems where the
lattice variables associated to the points $\odot$ are
not subject to  constraints except perhaps 
for conservation laws like total spin, charge, etc. 
Most of the Hamiltonians
in Condensed Matter or Stat. Mech. are of this form.
If the vertex Hamiltonian happens to have a continuous symmetry,
then the factorization of that symmetry would lead naturally
to an IRF model, whose renormalization can be studied using the
IRF-DMRG method.

The relation between the vertex-DMRG  and the 
IRF-DMRG algorithms, presented in  section III, is illustrated
diagrammatically in fig.17: the height $a$ (resp. $b) $ 
labels the different 
irreps (f.ex. the total  spins of the spin chains studied above)
that appear in the  tensor product of all the irreps contained in the 
block $B \;$ ( resp. $ B^R) $.
The intermediate height $j$ is obtained tensoring $a$ (resp. $b) $ 
with the irrep
carried by the vertex $\odot$,
which in the case of the spin chains  is a spin-s irrep. 
Finally, we must tensor $j \otimes j$ and pick up the identity 
irrep. 
Let us now find the analytic relation between the vertex
and IRF density matrices. We shall call ${\cal V}^B$ 
( resp. $ {\cal V}^{B^R}$ ) the Hilbert space 
associated to the block $B$ ( resp. $B^R)$. 
The tensor product decomposition (\ref{5.1}) becomes in this case

\begin{eqnarray}
& {\cal V}^B = \sum_a {\cal H}^B_a \otimes V_a & \label{5.15} \\
&{\cal V}^{B^R}= \sum_a   V_a  \otimes
 {\cal H}^{B^R}_a  & \nonumber
\end{eqnarray}

Using this decomposition the Hilbert space of the superblock
$B \odot \odot B^R$ becomes,

\begin{equation}
{\cal V}^{B \odot \odot B^R}
= \sum_{a,b} \; {\cal H}^B_a \otimes V_a \otimes V_s \otimes V_s
\otimes V_b \otimes {\cal H}^{B^R}_b
\label{5.16}
\end{equation}

The ground state of any rotational invariant Hamiltonian 
acting in this superblock can be written in the basis of
(\ref{5.16}) as follows (see fig.18),

\begin{eqnarray}
& |\psi_0> =
\sum \; | \xi_a \otimes m_a \otimes m_1 \otimes  m_2 \otimes m_b \otimes
\eta_b> 
 {\psi}_{\xi_a, \eta_b}(a,j,b) & \label{5.17} \\
& \left[ \begin{array}{lll}  j  & j & 0 \\
                           m_a + m_1 & m_b + m_2 & 0 \\
       \end{array} \right] \;
\left[ \begin{array}{lll}  a  & s & j \\
                           m_a & m_1 & m_a + m_1 \\
       \end{array} \right]
\left[ \begin{array}{lll}  s  & b & j \\
                           m_2 & m_b & m_2 + m_b \\
       \end{array} \right] & \nonumber 
\end{eqnarray}

\noindent
where $ {\psi}_{\xi_a, \eta_b}(a,j,b) $ is the IRF 
wave function of the ground state $|\psi_0>$.

The density matrix $\rho^{B \odot}$ can be obtained from (\ref{5.14}).
Using the properties of the CG coefficients we get,

\begin{eqnarray}
& \rho^{B \odot} = \sum
| \xi_a \otimes m_a \otimes m_1 > <  
\xi'_{a'} \otimes m'_{a'} \otimes m'_1 | & \label{5.18} \\
& = \frac{ \delta_{m_a+ m_1, m'_{a'} + m'_1 } }{2j +1} \;
{\rho}_{\xi_a}^{\xi'_{a'}} \left(
\begin{array}{ccc}   & a' &  \\
                    \ast &  & b \\
                     & a &  \end{array} \right) \;
\left[ \begin{array}{lll}  a  & s & j \\
                           m_a & m_1 & m_a + m_1 \\
       \end{array} \right]
\left[ \begin{array}{lll}  a'  & s & j \\
                           m'_{a'} & m'_1 & m'_{a'} + m'_1 \\
       \end{array} \right] & \nonumber
\end{eqnarray}

\noindent 
where

\begin{equation}
{\rho}_{\xi_a}^{\xi'_{a'}} \left(
\begin{array}{ccc}   & a' &  \\
                    \ast &  & j \\
                     & a &  \end{array} \right)
= \sum_{b, \eta_b} 
\psi_{\xi_a, \eta_b}(a,j,b) \; \psi^*_{\xi'_{a'}, \eta_b}(a',j,b) 
\label{5.19}
\end{equation}

\noindent 
coincides with the definition of the IRF-DM given in 
(\ref{3.5}). The relation (\ref{5.18})  between the vertex and IRF 
density matrices can be finally written as,

\begin{equation}
\rho^{B \odot} = \sum_j \frac{1}{2j +1} \; \rho^{S \bullet}
\label{5.20}
\end{equation}

The factor $1/(2j+1)$ takes care of the degeneracy of the irrep
$V_j$ in the CG decomposition $a \otimes s \rightarrow j$, and
guarantees the correct normalization conditions of both density matrices.

\section*{VII)  Conclusions and Perspectives}

We have generalized in this paper the DMRG method to
1d Hamiltonians of  IRF type and showed, in the examples
of the spin 1/2 SOS and RSOS models, that it gives very accurate
results. Our method is equivalent, by means of a vertex-IRF map, to the
standard DMRG method formulated by White for Hamiltonians of the
vertex type. This map  consists in the factorization of the 
symmetry group of the vertex theory. This factorization has
numerical and conceptual advantages. From a numerical point
of view one needs to keep a smaller number of states in the
IRF-DMRG  in order to achieve the same accuracy as in the
vertex-DMRG. The degeneracy of the eigenvalues of the vertex 
formulation, due to the symmetry, 
is absent in the IRF case, which makes the numerical
analysis more compact and stable. Conceptually the IRF-DMRG is also
very appealing since it employs tools and techniques well  known
in Statistical Mechanics,  
Integrable Systems, Multi-Matrix Algebras 
and Conformal Field Theory. Thus the
IRF states can be seen as paths of a Bratelli diagram, kinks
of a Theory of Solitons, discretized strings and  conformal
blocks in CFT. The formalism we have developed allow us to apply the DMRG
method to IRF states in a very natural way.

Let us mention some of the 
lines of research  which we believe deserve further study,

\begin{itemize}

\item  {\bf Higher Spin and Ferromagnetic Spin Chains:}  
In section V we 
have presented the necessary
tools to study  higher spin chains.
A particular interesting case is the  spin 1 chain , which has a rich
phase diagram. We shall show in a subsequent publication that
the string order parameter of den Nijs and Rommelse \cite{dNR},
which is used to characterize the Haldane phase, 
adopts a particular simple form when written in IRF variables.
In  fact  the IRF states constitute a
complete and orthonormal basis of valence bond states. 
In particular the  AKLT state \cite{AKLT}, which is a pure Haldane state,
is simply a straight path in the Bratelli diagram of the spin 1 
Heisenberg chain.

The IRF-DMRG is also very promissing for the study of ferromagnetic
systems, which seems to display a rich phase structure 
in the presence of magnetic fields \cite{OYA}.  
The vertex-DMRG method applied to ferromagnetic systems 
encounters the difficulty that  
the ground state  has a huge
degeneracy. As we have shown in this paper, the IRF-DMRG eliminates
this degeneracy, avoiding  the complications arisen from that fact.

\item {\bf t-J and Hubbard models:} The vertex-IRF map
can be straighforwardly applied to these models yielding
an IRF formulation where the spins form valence bonds. 
The IRF heights are now given by the couple 
(spin (j) ,charge (q)). In the case of Hubbard model the symmetry group 
is given by $SO(4)$ and it contains, in addition
to the  rotational group, the group of pseudo spin rotations. 
The factorization of
this larger group should reduce considerably the dimension of the
Hilbert Spaces.

\item {\bf Ladders:} These systems, which have received
considerably attention in the last 2 years, 
consist of a finite
number of coupled chains, with very interesting
properties ( for a review see \cite{DR}).
For spin ladders with a few number
of chains it is rather
simple to obtain their IRF version simply by taking the tensor product
of the irreps located on the rungs and performing
afterwards their  tensor
product along the chains. This procedure imitates the strong coupling
analysis applied to these kind of systems \cite{DR}. The IRF models
so obtained  has more than one link connecting different
heights, and so one has to generalize slightly the construction of this
paper. A similar multiplicity  phenomena  occurs in the theory of 
solitons \cite{GS2}. The IRF formulation of ladders could be useful to
clarify the relationship between their phases and those appearing in
higher spin chains.

\item {\bf Higher Dimensions} The DMRG philosophy is not confined to 1d, but
the standard DMRG  algorithms proposed so far
are one dimensional, despite of some 
2d applications to finite clusters  \cite{W2}.
The IRF-DMRG strengthen this point of view, since 
in particular the vertex-IRF map is a one-dimensional
operation. We should perhaps say that the vertex-IRF map
is really adimensional because the tensor product
operation  does not impose any particular
geometry or dimension. In connection with this problem it
may be useful to realize that the vertex-IRF map is a  
duality transformation similar to the Krammers-Wannier
duality or the Jordan-Wigner transformation. This interpretation
may serve  as a guide to construct  higher dimensional vertex-IRF maps.

\end{itemize}

There are still  many more 
topics to be considered in connection with the
DMRG. The DMRG method has arisen  as a numerical tool 
specially well adapted to 1d systems, but in our opinion 
its importance goes beyond its numerical success.
There are still some fundamental questions whose solution we would
like to know. It is perhaps not exagerate to say that 
new and radical  developments connected with the DMRG 
are likely to happen in the near future.

\vspace{3cm} {\bf Acknowledgements}

We would like to thank Steven White and Miguel A. Martin-Delgado
for conversations.

\vspace{2cm}{\em e-mail address}:
nishino@phys560.phys.kobe-u.ac.jp and 
sierra@sisifo.imaff.csic.es

\newpage

\section*{Figure Captions}

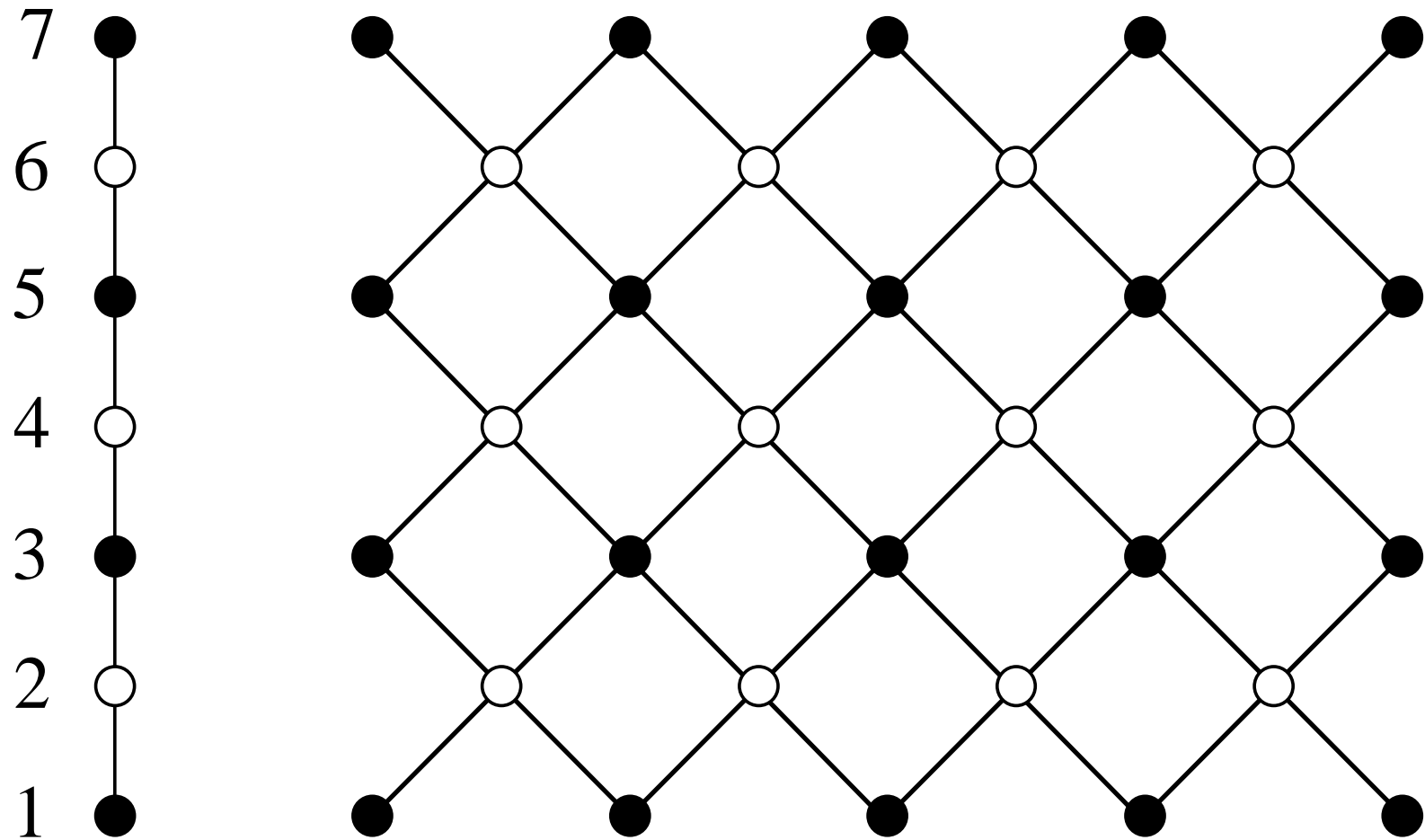
Fig.1.- Bratelli diagram associated to the Coxeter graph $A_7$.

\noindent
Fig.2.- In dark it is shown a path on the Bratelli diagram of 
Fig.1.

\noindent
Fig.3.- The string $S_{\ast,a}$ as a representative of the class
of all paths on a Bratelli diagram starting at $\ast$ and ending at
$a$.

\noindent
Fig.4.- A string $S_{\ast,a}$ absorbs a point $\bullet$ 
which carries an  allowed  state $b$, becoming a new string $S'_{\ast,b}$.

\noindent
Fig.5.- Diagrammatic representation of the string operators
(\ref{2.7}).

\noindent
Fig.6.- Top: diagrammatic representation of the equation (\ref{2.8}).
Bottom: Diagrammatic reconstruction of the Hamiltonian $H^{S,3 \bullet}$.

\noindent
Fig.7.- Diagrams of the $T$ and $T^\dagger$ operators
(\ref{2.10}) ( T for truncation and for triangle).

\noindent
Fig.8.- The normalization condition (\ref{2.13}) interpreted as a kind of
annihilation process of triangles.

\noindent
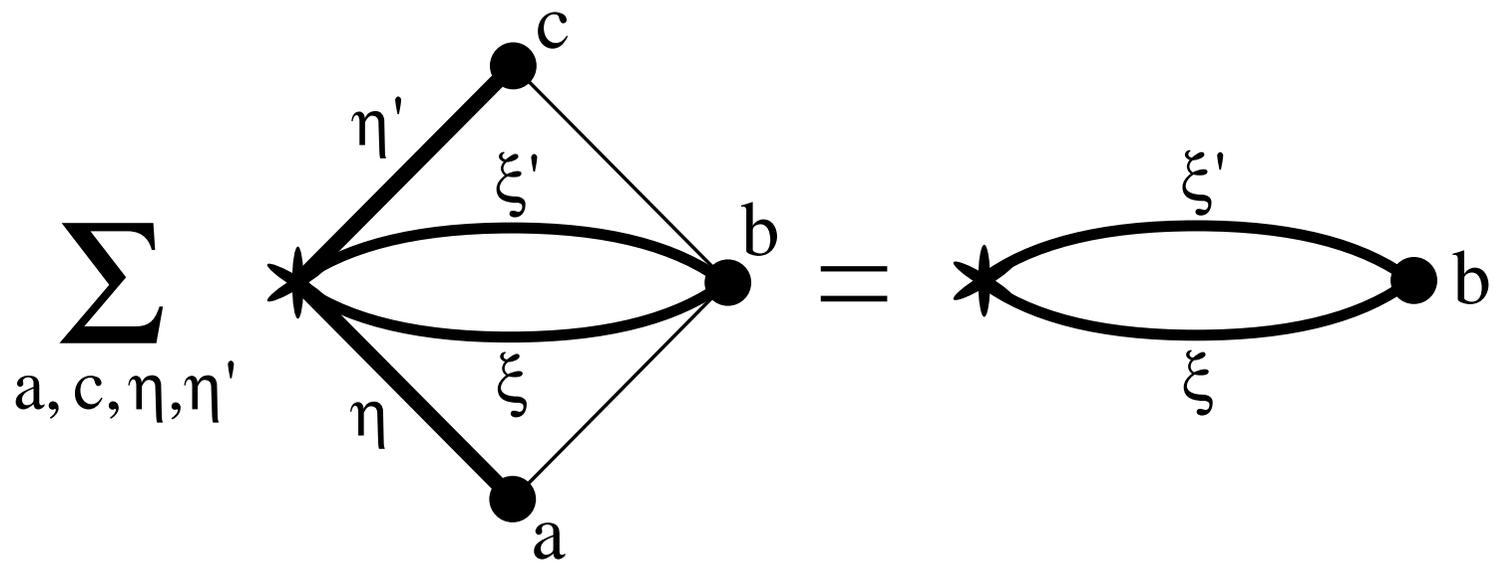
Fig.9.- Eq.(\ref{2.16}) in pictorical form.

\noindent
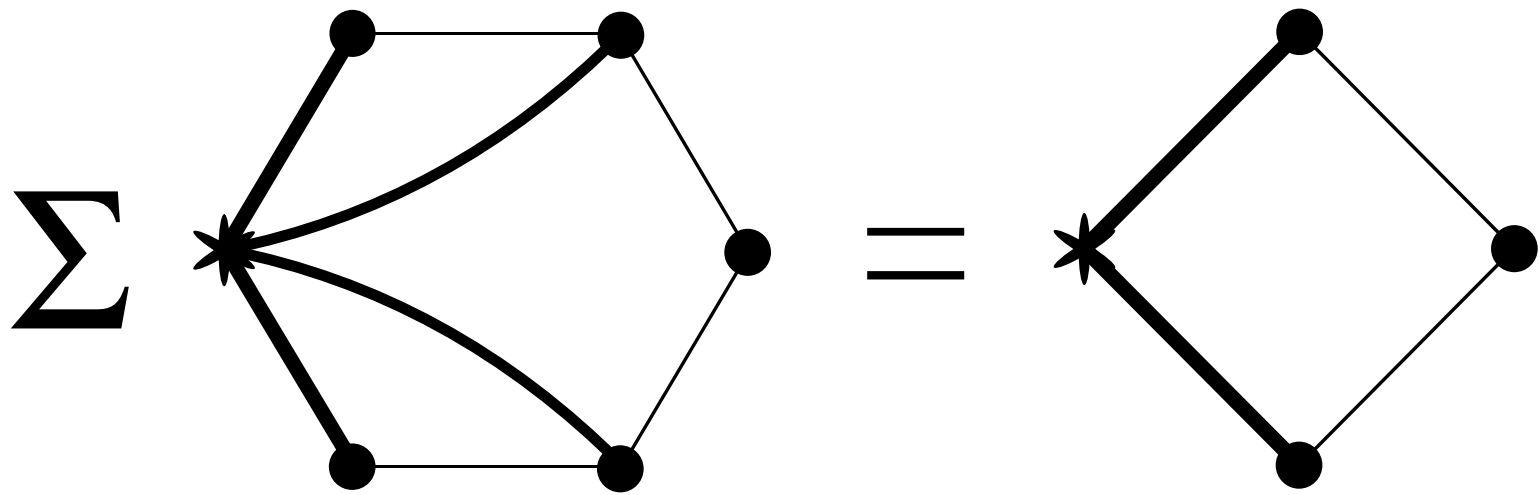
Fig.10.- The renormalization of ${\cal O}^{S, 2\bullet}$ ( 
see eq.(\ref{2.17})). From figs.8.9 and 10 we see that
the RG procedure is a kind of sewing or gluing construction 
involving triangles, plaquettes and higher n-gons. 

\noindent
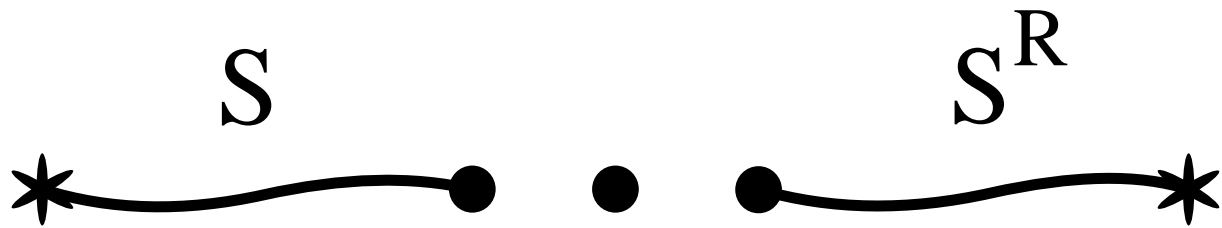
Fig.11.- The "super-string" configuration that leads to the
infinite system  IRF-DMRG algorithm.

\noindent
Fig.12.- Coxeter graph $A_\infty$.

\noindent
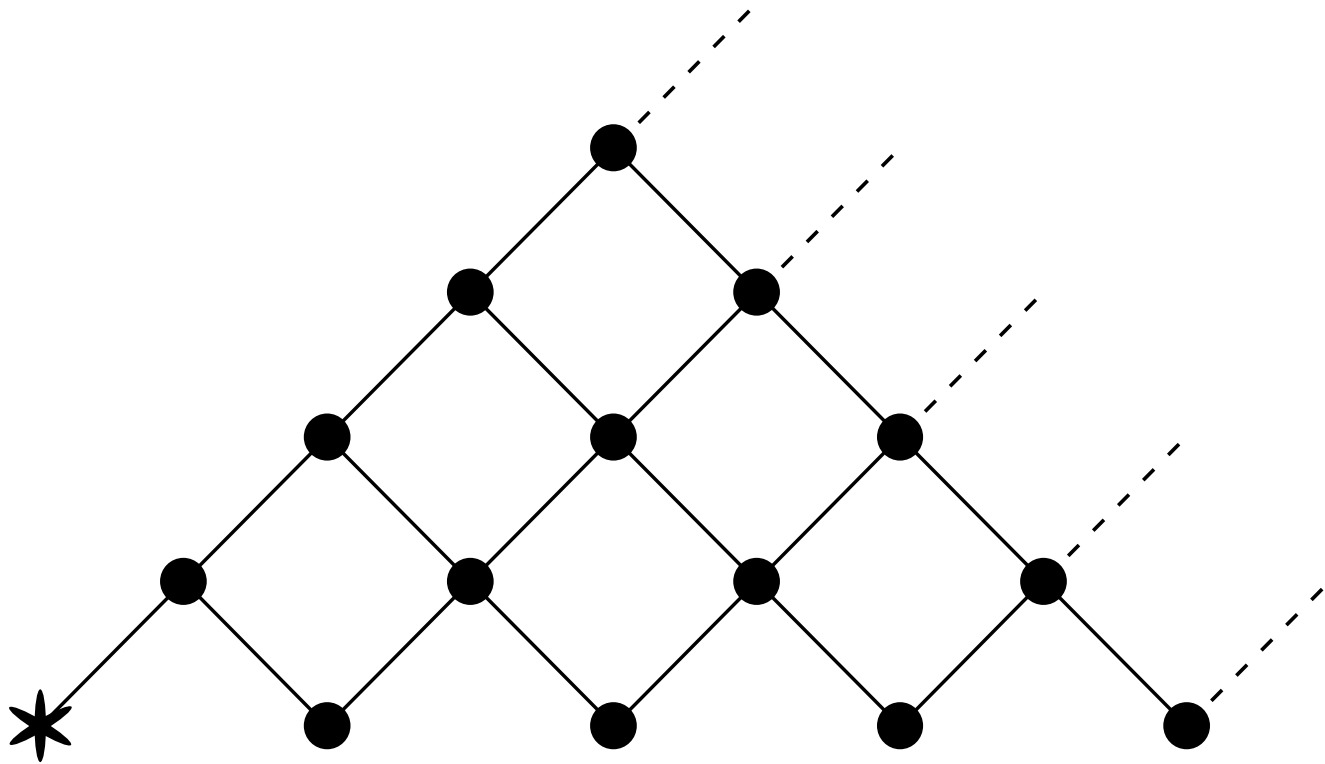
Fig.13.- Bratelli diagram built up using the Coxeter graph
$A_\infty$.

\noindent
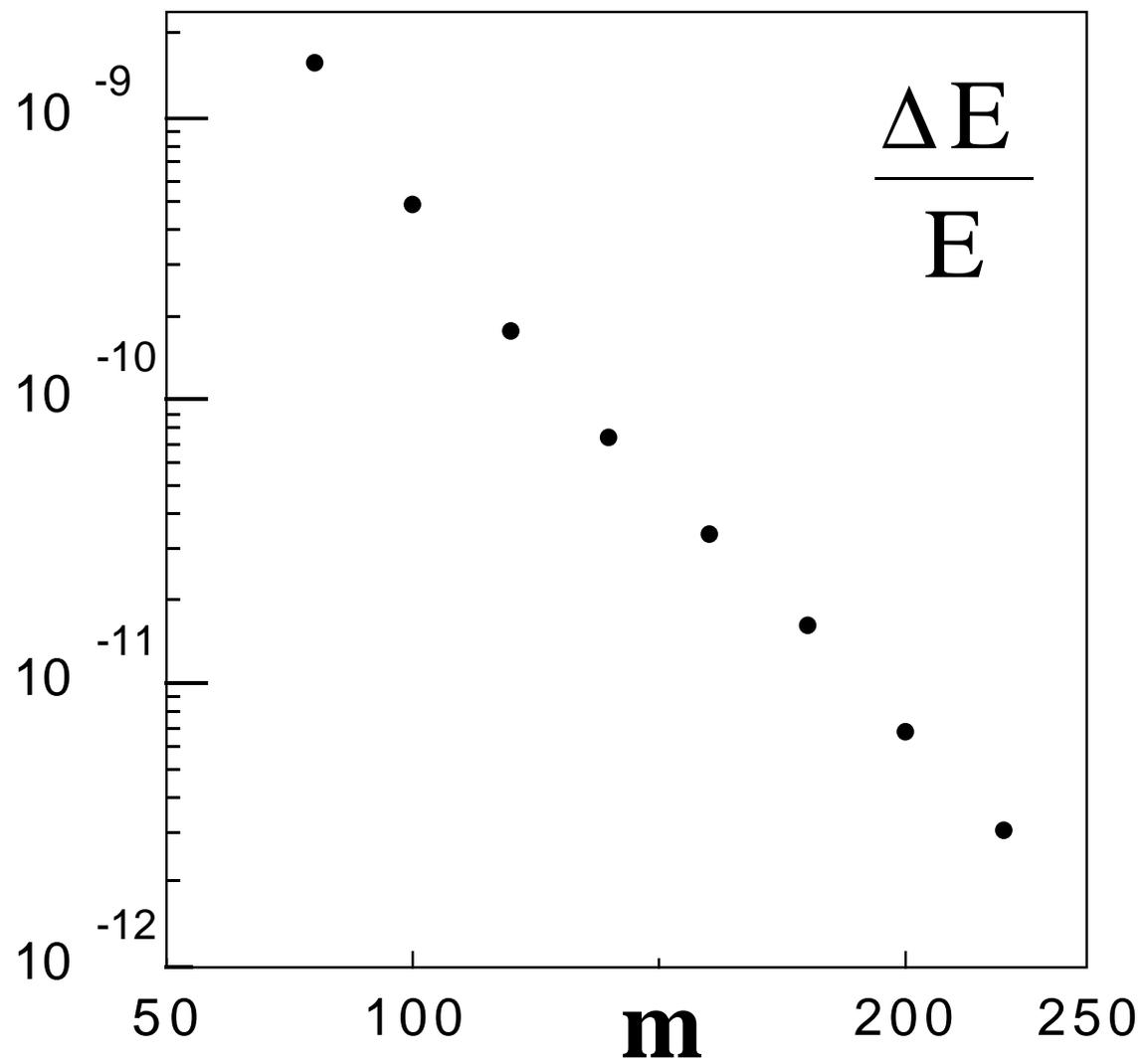
Fig.14.- Plot of the deviation of the IRF-DMRG ground state
energy of a s=1/2 chain with 512 sites, as a function of the
number of states retained m.

\noindent
Fig.15.- Coxeter garph $A_r$.

\noindent
Fig.16.- Graphical representation of the vertex-IRF map.  
Notice that the IRF points $\bullet$ and the vertex points
$\odot$ belong to  lattices which are dual one another.
Indeed the vertex-IRF map is a kind of duality transformation.

\noindent
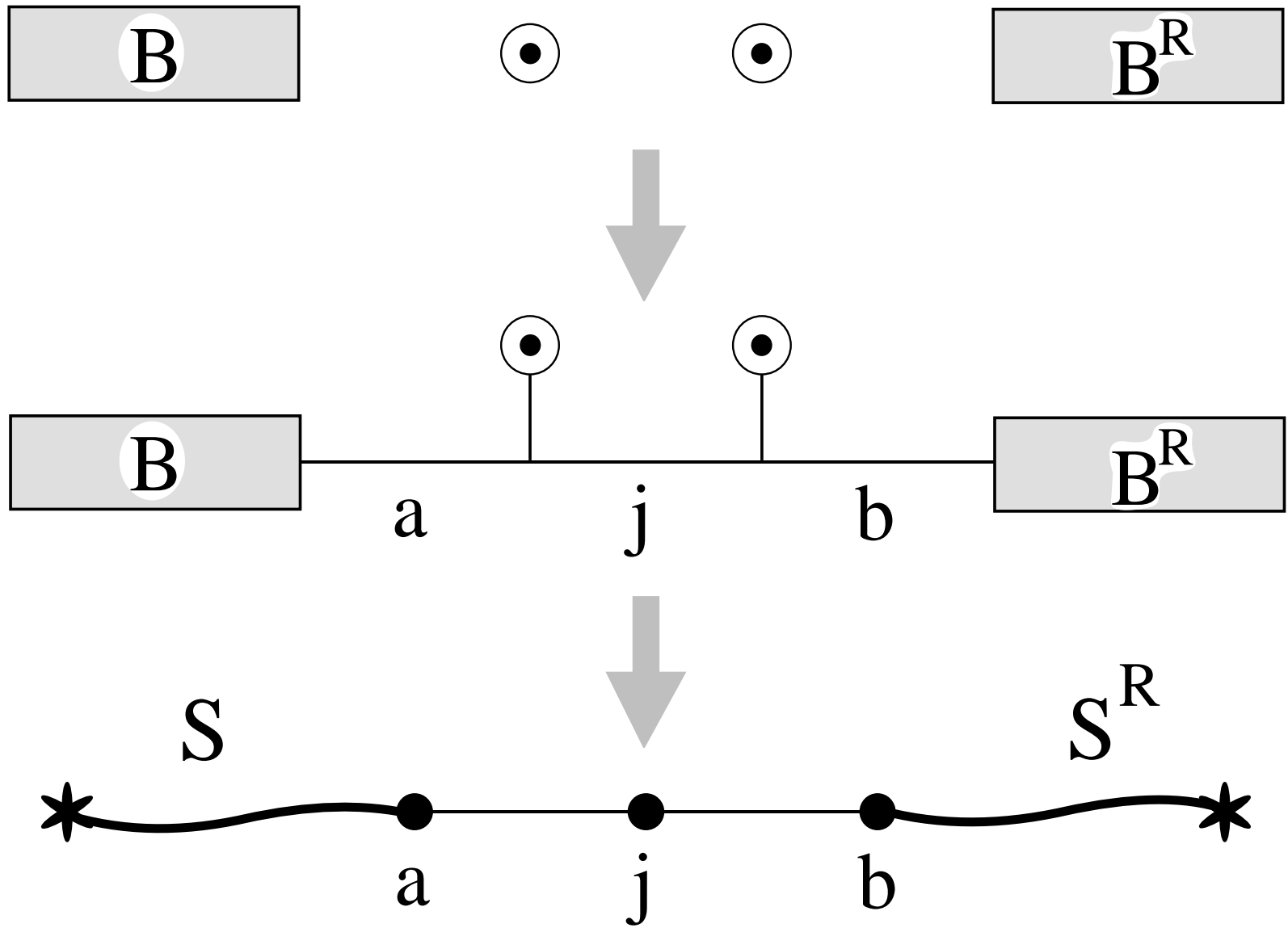
Fig.17.- The vertex-IRF map that relates the vertex-DMRG  and 
the IRF-DMRG algorithms.

\noindent
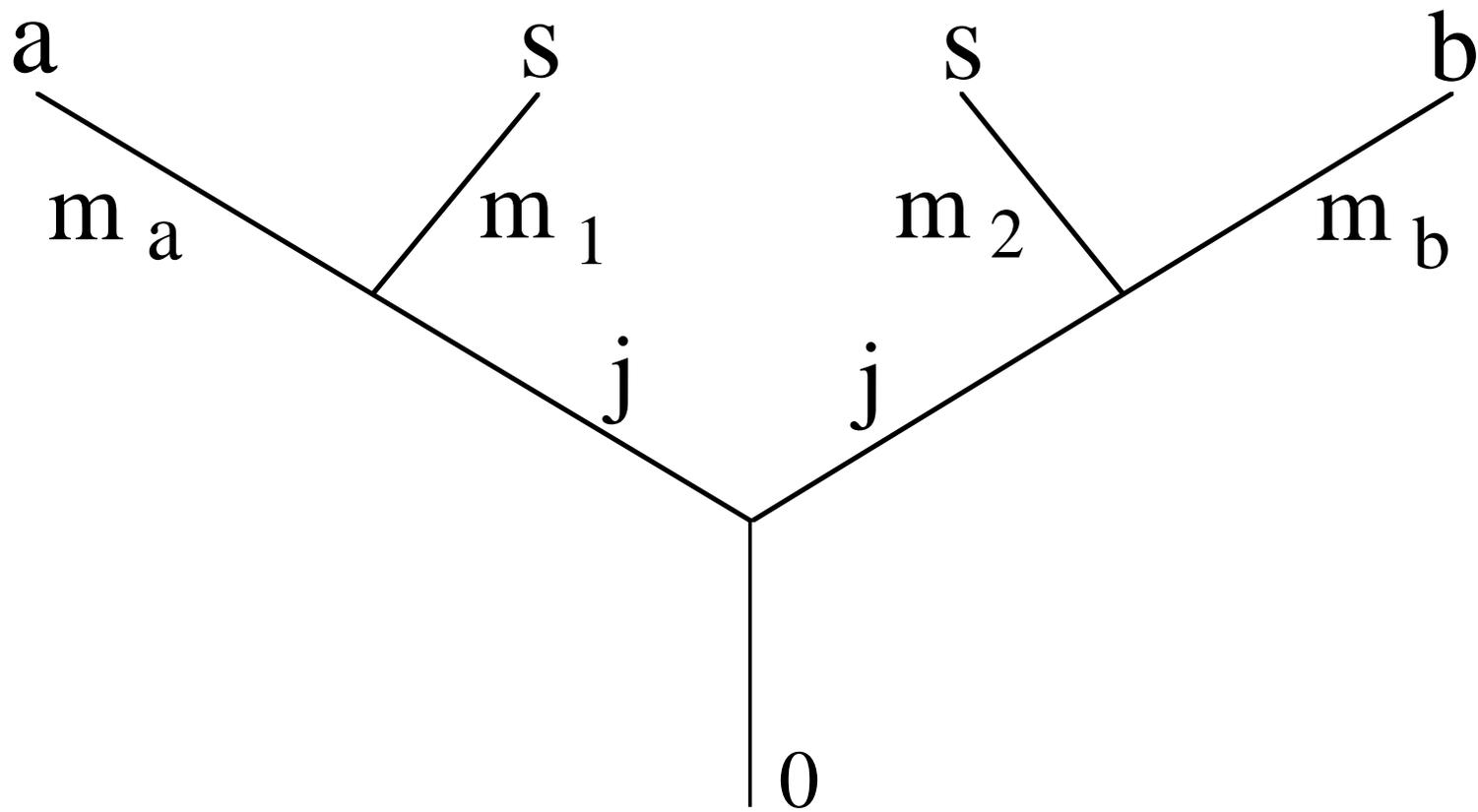
Fig.18.- Here we show the CG decompositions involved in fig.17.


\newpage



\begin{thebibliography}{99}

\bibitem{W1} S.R. White, Phys. Rev. Lett. 69, 2863 (1992);
Phys. Rev. B 48, 10345 (1993).                    

\bibitem{Wi} K. G. Wilson, Rev. Mod. Phys. 47, 773 (1975).


\bibitem{SLAC} S.D. Drell, M. Weinstein and S. Yankielowicz,
Phys. Rev. D 16, 1769 (1977)

\bibitem{Paris} R. Jullien, P. Pfeuty, J.N. Fields and S. Doniach,
Phys. Rev. B 18, 3568 (1978)

\bibitem{BRG} P. Pfeuty, R. Jullien and K.A. Penson, in:  Real
Space Renormalization, eds. T.W. Burkhardt and J.M.J. van Leeuwen,
Series Topics in Current Physics 30, Springer-Verlag, 1982.


\bibitem{hT} J. Gonzalez, M.A. Martin-Delgado, M.A.H. Vozmediano,
Quantum Electron Liquids and Hight T$_c$ Superconductivity,
Lecture Notes in Physics m38, Springer-Verlag 1995. For the
RG consult chapter 11.


\bibitem{WH} S.R. White and D.A. Huse, Phys. Rev. B 48, 3844 (1993)


\bibitem{SA} E.S. Sorensen and I. Affleck, Phys. Rev. Lett. 71, 
1633 (1993); Phys. Rev. B 49, 15771 (1994).


\bibitem{WNS} S.R. White, R.M. Noack and D.J. Scalapino,
Phys. Rev. Lett. 73, 886 (1994).


\bibitem{W2} S.R. White, cond-mat/9604129.


\bibitem{NO} T. Nishino and K. Okunishi, J. Phys. Soc. Jpn. 65,
891 (1996).


\bibitem{OR} S. \"Ostlund and S. Rommer,  Phys. Rev. Lett. 75, 3537
(1995); cond-mat/9606213.


\bibitem{X} T. Xiang, Phys. Rev. 53 , 10445 (1996).


\bibitem{Ni} T. Nishino, J. Phys. Soc. Jpn. 64, 3598 ( 1995).

\bibitem{X2} R.J. Bursill, T. Xiang and G.A. Gehring, cond-mat/9609001.
 


\bibitem{MS1} M.A. Martin-Delgado and G. Sierra,
Int. J. Mod. Phys. A 11, 3145 (1996).

\bibitem{qRG}   M.A. Martin-Delgado and G. Sierra, Phys. Rev. Lett.
76, 1146 (1996).


\bibitem{Role}  M.A. Martin-Delgado and G. Sierra, Phys. Lett. B 364,
41 (1995).


\bibitem{CBRG} M.A. Martin-Delgado, J. Rodriguez-Laguna  and G. Sierra,
Nucl. Phys. B 473 [FS], 685 (1996).


\bibitem{B} R.J. Baxter, "Exactly Solved Models in Statistical
Mechanics", Academic Press, London, 1982.


\bibitem{ABF}G.E. Andrews, R.J. Baxter and P.J. Forrester,
J. Stat. Phys. 35, 193 (1984).


\bibitem{Affine} E. Date, M.Jimbo, T. Miwa and M.Okado,
Phys. Rev. B 35, 2105 (1987). 



\bibitem{GHJ} F.M. Goodman, P.M. de la Harpe and V. F. R. Jones,
"Coxeter-Dynkin Diagrams and Towers of Algebras" , MSRI 
Publications/Springer-Verlag, New York (1989).


\bibitem{GS1} C. Gomez and G.Sierra, Intern. J. Mod. Phys. A6, 2045 (1991). 



\bibitem{P1} V. Pasquier, Nucl. Phys. B285 [FS19], 162 (1987)
; J. Phys. A 20, L217, L221 (1987).


\bibitem{LW} E.H.Lieb and F.Y. Wu, in Phase transitions and critical
phenomena, vol. I, ed. C. Domb and M.S. Green, Academic Press (1972).



\bibitem{B2} R.J. Baxter, Ann. Phys. 76, 25 (1973).


\bibitem{Ji2} M. Jimbo, T. Miwa and M.Okado, 
Comm. Math. Phys. 116, 507 ( 1988).


\bibitem{P2} V. Pasquier, Comm. Math. Phys. 118, 355 (1988).



\bibitem{BPZ} A.A. Belavin, A. M. Polyakov and A.B. Zamolodchikov,
Nucl. Phys. B 241, 333, (1984).



\bibitem{qb} C. Gomez, M. Ruiz-Altaba and G. Sierra, " Quantum Groups in
Two Dimensional Physics", Cambridge University Press, 1996.


\bibitem{O} A. Ocneanu, "Quantized groups, string algebras
and Galois theory of algebras", London Math. Soc. Lecture Notes
136, 119 ( 1989).



\bibitem{Zamos} A.B. Zamolodchikov and A.Bl. Zamolodchikov,
Ann. Phys. 120, 253 (1979).


\bibitem{WN} S.R. White and R.M. Noack, Phys. Rev. Lett. 68, 3487 (1992).




\bibitem{ABBBQ} F.C. Alcaraz, M.N. Barber and M.T. Batchelor,
R.J. Baxter and G.R.W. Quispel, J. Phys. A: Math. Gen. 20, 6397 (1987).


\bibitem{H} C.J. Hamer, J. Phys. A: Math Gen, 19, 3335 (1986).


\bibitem{ABB} F.C. Alcaraz, M.N. Barber and M.T. Batchelor,
Phys. Rev. Lett. 58, 771 (1987).



\bibitem{PS} V. Pasquier and H. Saleur, Nucl. Phys. B 330, 523 (1990).


\bibitem{TL} H.N.V. Temperley and E. Lieb, Proc. Roy. Soc. London
A 322, 251 (1971).


\bibitem{Cardy} H.W.J. Blote, J.L. Cardy and M.P. Nightingale,
Phys. Rev. Lett. 56, 742 (1986).


\bibitem{Aff} I. Affleck, Phys. Rev. Lett. 56, 746 (1986).


\bibitem{van} J.M. Van der Broeck and L.W. Schwartz,
SIAM J. Math. Anal. 10, 658 (1979). 
 

\bibitem{dNR} M. den Nijs and K. Rommelse, Phys. Rev. B 40, 4709 (1989)

\bibitem{AKLT} I. Affleck, T. Kennedy, E. Lieb and H. Tasaki,
Commun. Math. Phys. 115, 477 (1988).


\bibitem{OYA} M. Osaka, M. Yamanaka and I. Affleck, cond-mat/9610168.


\bibitem{DR} E. Dagotto and T.M. Rice, Science 271, 618 (1996).


\bibitem{GS2} C. Gomez and G. Sierra, Nucl. Phys. B 419 [FS], 589
(1994).





\end{thebibliography}
\end{document}